%% file: main.tex
\documentclass[onefignum,onetabnum]{siamonline190516}

\input{info_shared}

\ifpdf
\hypersetup{
  pdftitle={Bayesian UQ using Deep Neural Networks},
  pdfauthor={S. Lan, S. Li, and B. Shahbaba}
}
\fi

\begin{document}

\maketitle

\begin{abstract}
Due to the importance of uncertainty quantification (UQ), Bayesian approach to inverse problems has recently gained popularity in applied mathematics, physics, and engineering. However, traditional Bayesian inference methods based on Markov Chain Monte Carlo (MCMC) tend to be computationally intensive and inefficient for such high dimensional problems. To address this issue, several methods based on surrogate models have been proposed to speed up the inference process. More specifically, the calibration-emulation-sampling (CES) scheme has been proven to be successful in large dimensional UQ problems.  
In this work, we propose a novel CES approach for Bayesian inference based on deep neural network models for the emulation phase. The resulting algorithm is computationally more efficient and more robust against variations in the training set. Further, by using an autoencoder (AE) for dimension reduction, we have been able to speed up our Bayesian inference method up to three orders of magnitude. Overall, our method, henceforth called \emph{Dimension-Reduced Emulative Autoencoder Monte Carlo (DREAMC)} algorithm, is able to scale Bayesian UQ up to thousands of dimensions for inverse problems. Using
two low-dimensional (linear and nonlinear) inverse problems we illustrate the validity of this approach. Next, we apply our method to two high-dimensional numerical examples (elliptic and advection-diffussion) to demonstrate its computational advantages over existing algorithms.
\end{abstract}



\section{Introduction}
There is a growing interest in uncertainty quantification (UQ) in the field of applied mathematics and its applications in physical sciences, biological sciences, and engineering, where UQ is commonly used to calibrate model inadequacy, carry out sensitivity analysis, or solve optimal control problems under uncertainty. As a result, Bayesian methods for inverse problems (e.g., reservoir modeling, weather forecasting) have become increasingly popular. 
Models in these application domains are usually constrained to physical or biological laws and are typically represented as ordinary or partial differential equation (ODE/PDE) systems.
Implementing Bayesian UQ for such inverse problems is quite difficult because they involve computationally intensive simulations for 1) solving the ODE/PDE systems, and 2) sampling from the resulting high dimensional posterior distributions. To address these issues, we follow the work of \cite{cleary2020} and propose a more scalable framework for Bayesian UQ that combines ensemble Kalman methods and infinite-dimensional Markov Chain Monte Carlo (MCMC) algorithms.

Ensemble Kalman (EnK) methods, originated from geophysics \cite{evensen1994}, have achieved significant success in state estimation for complex dynamical systems with noisy observations \cite{evensen1996,houtekamer2001,aanonsen2009,Evensen2009,Evensen03a,Kalnay_2002,Law2015,Houtekamer_2016,Jurek_2021}.
More recently, these methods have been used to solve inverse problems with the objective of estimating the model parameters instead of the states \cite{Oliver_2008,Chen_2011,Emerick_2013,Iglesias_2013,Iglesias_2016,Evensen_2018,Katzfuss_2019,Garbuno-Inigo_2020}.
As a gradient-free optimization algorithm based on a small number of ensembles, these methods gained increasing popularity for solving inverse problems since they can be implemented non-intrusively and in parallel.
However, due to the collapse of ensembles \cite{Schillings_2017a,Schillings_2017b,deWiljes2018,chada2019}, they tend to underestimate the posterior variance and thus fail to provide a rigorous basis for systematic UQ. 
To alleviate this issue, Cleary et al.~combine Kalman methods with MCMC in three steps \cite{cleary2020}: (i) calibrate models with ensemble Kalman methods, (ii) emulate the parameter-to-data map using evaluations of forward models, and (iii) generate posterior samples using MCMC based on cheaper emulators. The resulting approach is called \emph{Calibration-Emulation-Sampling (CES)}.
Two immediate benefits of such a framework are: 1) reusing expensive forward evaluations, and 2) computationally efficient surrogates in the MCMC procedure.

For the emulation component, \cite{cleary2020} rely on Gaussian Process (GP) models, which have been widely used for emulating computer models \cite{Sacks_1989}, uncertainty analysis \cite{Oakley2002}, sensitivity analysis \cite{Oakley2004}, and computer code calibration \cite{Kennedy2001,Higdon_2004,OHAGAN2006}. While not adopted in \cite{cleary2020}, the derivative information can be extracted from GP \cite{stephenson10,lan2016} to improve the sampling component \cite{Kennedy2001,Oakley2002}.
In general, however, GP emulation involves intensive computation ($\mathcal O(N^3)$ with $N$ input/training data pairs) and does not scale well to high dimensions. Additionally, the prediction accuracy of GP emulator highly depends on the training set, which usually demands a substantial effort through careful experimental design \cite{Sacks_1989,santner03}. For these reasons, we propose to use deep neural network (NN) \cite{Goodfellow2016} for the emulation component of CES.

A deep artificial NN consists of multiple layers mapping the input (predictors) to the output (response). Each layer contains neurons (units) and synapses (connections) defining the architecture of NN. The resulting map also depends on a set of weights, biases, and activation functions, which are respectively analogous to regression coefficients, intercepts, and link functions in generalized linear models. Composition of multiple layers in different types makes the whole network capable of learning complex non-linear relationships, hence making NN a flexible functional approximation tool \cite{Bengio_2009,LeCun_2015,SCHMIDHUBER_2015}.
Deep NN has achieved enormous successes in image processing \cite{Krizhevsky_2012,CIRESAN_2012}, speech recognition \cite{Graves_2004}, natural language processing \cite{Gers_2001,Mesnil_2015}, bioinformatics \cite{Chicco_2014,Senior_2020}, and many other areas \cite{Silver_2016,Silver_2017}.
Using stochastic batch optimization algorithms such as stochastic gradient descent \cite{Bottou_1999}, the computational cost of NN can be reduced to $\mathcal O(Np)$ with $N$ being the size of the training data, and $p$ being the total number of NN parameters \cite{Bottou_2004}.

In this paper, we particularly focus on convolutional neural networks (CNN), which are commonly used in image recognition \cite{Krizhevsky_2012}. In inverse problems, parameter functions (i.e., functions treated as parameters) are defined on finite 2d or 3d domains that can be viewed as images. Therefore, we expect CNN to be more suitable than regular dense neural networks (DNN) for such problems. This novel observation could lead to more effective and computationally efficient emulators.
Although there have been a few related works where CNN is used on actual images \cite{Jin_2017,McCann_2017,Wang_2020} or as a prior for spatiotemporal dynamics \cite{ZAMMITMANGION2020} in inverse problems,
to the best of our knowledge, this is the first application of CNN for training generic emulators in Bayesian inverse problems.

Besides computational challenges associated with building emulators, sampling from posterior distributions in inverse problems is also a challenging task due to the high dimensionality of the target distribution.
Traditional Metropolis-Hastings algorithms defined on the finite-dimensional space suffer from deteriorating mixing times upon mesh-refinement or increasing dimensions \cite{roberts97,Roberts_1998,Beskos_2009}.
To overcome this deficiency, a new class of `dimension-independent' MCMC methods \cite{beskos08, beskos11, cotter13, law2014, beskos14, beskos2017} has been proposed on infinite-dimensional Hilbert space.
Despite of the robustness to increasing dimensions, these $\infty$-dimensional MCMC algorithms are still computationally demanding.
Several recent papers have attempted to address this issue by using dimension reduction methods to find an intrinsic low-dimensional subspace that contains the most amount of information on the posterior distribution \cite{cui16, constantine2016, LAN2019a}.
In this paper, we adopt autoencoder (AE) \cite{Hinton_2006b} to learn a low-dimensional latent representation of the parameter space with an encoder (original $\to$ latent) and a decoder (latent $\to$ reconstruction). 

Combining CNN and AE, we propose a class of hybrid MCMC algorithms called \emph{Dimension Reduced Emulative Autoencoder Monte Carlo (DREAMC)}, which can improve and scale up the application of the CES framework for Bayesian UQ from hundreds of dimensions (with GP emulation) \cite{cleary2020} to thousands of dimensions (with NN emulation). In summary, this paper makes multiple contributions as follows:
\begin{enumerate}[nosep]
\item apply CNN to train emulators for Bayesian inverse problems;
\item embed AE in CES to significantly improve its computational efficiency;
\item scale Bayesian UQ for inverse problems up to thousands of dimensions with DREAMC.
\end{enumerate}

The rest of the paper is organized as follows. Section \ref{sec:BUQ} provides a brief overview of Bayesian inverse problems and the CES framework for UQ. Further, gradient based MCMC algorithms ($\infty$-MALA and $\infty$-HMC, not adopted in \cite{cleary2020}) are reviewed for sampling. In Section \ref{sec:scaleBUQ}, we apply various neural networks, including DNN, CNN, and AE, to scale up Bayesian UQ for inverse problems. Details of emulating functions, extracting gradients and reducing dimensions will also be discussed. Then, we illustrate the validity of DREAMC in Section \ref{sec:illust} with simulated linear and nonlinear inverse problems. In Section \ref{sec:numerics}, using two high-dimensional inverse problems involving an elliptic equation and an advection-diffusion equation, we demonstrate that our methods can speed up traditional MCMC algorithms up to three orders of magnitude. Section \ref{sec:con} concludes with some discussions on future research directions.

\section{Bayesian UQ for Inverse Problems: Calibration-Emulation-Sampling}\label{sec:BUQ}

In many PDE-constrained inverse problems, we are interested in finding an unknown function, $u$ (symbolizing an `unknown' function), given the observed data, $y$.
The function $u$ usually appears as a parameter (hence termed `parameter function' thereafter) in the inverse problem.
For example, $u$ could be the (log)-transmissivity of subsurface flow that appears as a coefficient function in an elliptic PDE (Section \ref{sec:elliptic}) or the initial condition of a time-dependent advection-diffusion problem (Section \ref{sec:adif}). Let $\mbX$ and $\mbY$ be two separable Hilbert spaces. (This is assumed for the convenience of developing probability spaces and can be relaxed to separable subspaces \cite{dashti2017,stuart10}.)
There is a forward parameter-to-observation mapping $\mG:\mbX \rightarrow \mbY, \; u\mapsto \mG(u)$ from the parameter space $\mbX$ to the data space $\mbY$ (e.g. $\mbY=\mathbb R^m$ for $m\geq 1$) that connects $u$ to $y$ as follows:
\begin{equation}\label{eq:forward}
y=\mG(u) + \eta, \qquad \eta\sim \mN(0,\Gamma)
\end{equation}
We can then define the potential function (negative log-likelihood), $\Phi:\mbX\times \mbY\to \mathbb R$, as follows:
\begin{equation}\label{eq:gauss_nz}
\Phi(u;y) = \frac{1}{2} \Vert y-\mG(u)\Vert^{2}_{\Gamma} = \frac{1}{2} \langle y-\mG(u), \Gamma^{-1} (y-\mG(u)) \rangle
\end{equation}
The forward mapping $\mG$ represents physical laws usually expressed as large and complex ODE/PDE systems and could be highly non-linear. Therefore, repeated evaluations of the likelihood (and $\mG(u)$) could be computationally demanding for different values of $u$.

In the Bayesian setting, a prior measure $\mu_0$ is imposed on $u$, independent of $\eta$. 
For example, we could assume a Gaussian prior $\mu_0 = \mathcal N(0,\mathcal C)$ with the covariance $\mathcal C$ being a positive, self-adjoint and trace-class (a.k.a. nuclear) operator on $\mbX$. (A trace-class operator $\mathcal C$ has summable eigenvalues $\lambda_n(\mC)$, i.e. $\tr(\mC):=\sum_n\lambda_n(\mC)<\infty$.)
Then we can obtain the posterior of $u$, denoted as $\mu_{u|y}$, using Bayes' theorem \cite{stuart10,dashti2017}: 
\begin{equation*}
\frac{d\mu_{u|y}}{d\mu_0}(u) = \frac{1}{Z}\,\exp(-\Phi(u;y)) \ , \quad \textrm{if} \ 0< Z:=\int_{\mbX} \exp(-\Phi(u;y)) \mu_0(du) < +\infty \ .
\end{equation*}
For simplicity, we drop $y$ and denote the posterior distribution and potential function as $\mu(du)$ and $\Phi(u)$ respectively.
The Bayesian inverse problems involve estimating $u$ and quantifying the associated uncertainty. For example, we are interested in tracing back the initial condition based on down-stream observations in an advection-diffusion problem (Section \ref{sec:adif}). This reduces to learning the posterior distribution $\mu(du)$, which can exhibit strong non-Gaussian behavior, posing enormous difficulties for commonly used inference methods such as MCMC.

In addition to the above-mentioned challenges in Bayesian UQ for inverse problems, the high dimensionality of the discretization of $u$ makes the forward evaluation computationally intensive and imposes challenges on the robustness of sampling algorithms.
To this end, the CES framework has been recently proposed for approximate Bayesian parameter learning. It consists of the following three stages \cite{cleary2020}:
\begin{enumerate}[itemsep=0pt]
\item {\bf Calibration}: using optimization-based algorithms (ensemble Kalman) to obtain parameter estimation and collect expensive forward evaluations for the emulation step;
\item {\bf Emulation}: recycling forward evaluations in the calibration stage to build an emulator for sampling;
\item {\bf Sampling}: sampling the posterior approximately based on the emulator, which is much cheaper than the original forward mapping.
\end{enumerate}

The CES scheme is promising for high-dimensional Bayesian UQ in inverse problems. Emulation bypasses the expensive evaluation of original forward models (dominated by the cost of repeated forward solving of ODE/PDE systems) and reduces the cost of sampling to a small computational overhead. The sampling also benefits from the calibration, which provides MCMC algorithms with a good initial point in the high density region so that the burning time can be reduced.

In this paper, we choose NN for emulation instead of GP (used in \cite{cleary2020}) for the computational efficiency and flexibility. Moreover, we extract the gradient evaluations directly from NN to implement gradient-based $\infty$-dimensional MCMC for sampling. We also adopt AE as a dimension reduction technique to further improve the efficiency of the original CES method \cite{cleary2020}. 
In the following, we review ensemble Kalman methods for calibration and $\infty$-dimensional MCMC algorithms for sampling.

\subsection{Calibration -- Ensemble Kalman (EnK) Methods}
For state-space models, Kalman filter \cite{Kalman_1960} and its ensemble variants \cite{evensen1994,Evensen_1996} have become standard solvers because of their linear computational complexity.
Recently, EnK methods have been introduced to solve inverse problems with the objective of estimating parameters rather than states \cite{Chen_2011,Iglesias_2013,Iglesias_2016,Evensen_2018,Garbuno-Inigo_2020}.

Initializing $J$ ensemble particles $\{u^{(j)}\}_{j=1}^J$ with, for example, prior samples,
the basic ensemble Kalman inversion (EKI) method 
evolves each ensemble according to the following equation \cite{Schillings_2017a}:
\begin{equation}\label{eq:enkf_cont}
\frac{d u^{(j)}}{dt} = \frac1J \sum_{k=1}^J \left\langle \mG(u^{(k)}) - \bar\mG, y - \mG(u^{(j)}) + \sqrt{\Sigma} \frac{dW^{(j)}}{dt} \right\rangle_{\Gamma} (u^{(k)} - \bar u)
\end{equation}
where $\bar u :=\frac1J\sum_{j=1}^J u^{(j)}$, $\bar \mG:=\frac1J\sum_{j=1}^J \mG(u^{(j)})$, and $\{W^{(j)}\}$ are independent cylindrical Brownian motions on $\mbY$. We can set $\Sigma=0$ to remove noise for an optimization algorithm, or $\Sigma=\Gamma$ to add noise for dynamics that transform the prior to the posterior in one time unit for linear forward maps \cite{Schillings_2017a,Garbuno-Inigo_2020}.

A variant of EKI for approximate sampling from the posterior $\mu(du)$ is an ensemble Kalman sampler (EKS) \cite{Garbuno-Inigo_2020,Garbuno_Inigo_2020}.
This is obtained by adding a prior-related damping term as in \cite{chada2019}, and modifying the position-dependent noise in Equation \eqref{eq:enkf_cont}:
\begin{equation}\label{eq:ens_cont}
\frac{d u^{(j)}}{dt} = \frac1J \sum_{k=1}^J \left\langle \mG(u^{(k)}) - \bar\mG, y - \mG(u^{(j)}) \right\rangle_{\Gamma} (u^{(k)} - \bar u) - C(u) \mC^{-1} u^{(j)} + \sqrt{2C(u)} \frac{dW^{(j)}}{dt} 
\end{equation}
with $C(u):= \frac1J \sum_{j=1}^J  (u^{(j)} - \bar u) \otimes (u^{(j)} - \bar u)$.

\begin{figure}[t]
  \begin{center}
     \includegraphics[width=1\textwidth,height=.3\textwidth]{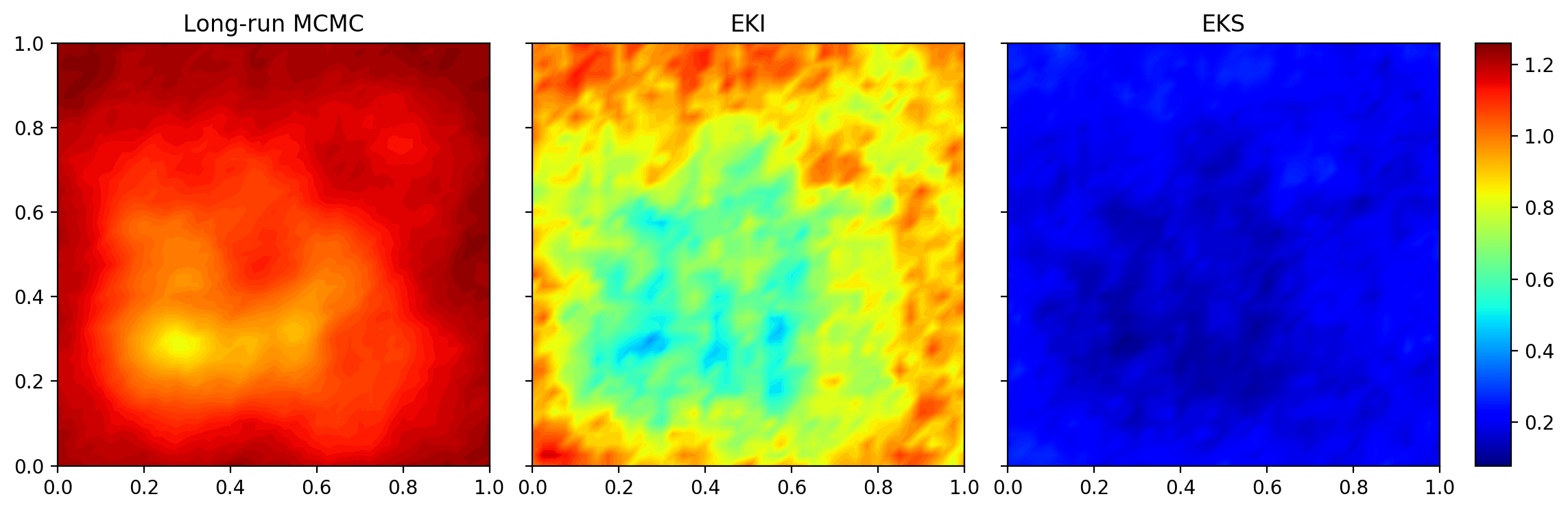}
  \end{center}
  \caption{Comparing the estimation of standard deviation by MCMC (left panel) and ensemble Kalman methods (middle and right panels) with 100 ensembles in an elliptic inverse problem (Section \ref{sec:elliptic}).}
  \label{fig:ensbl_std}
\end{figure}

Both EKI and EKS are implemented by discretizing \eqref{eq:enkf_cont} and \eqref{eq:ens_cont} respectively and updating $u^{(j)}_n$ iteratively for $n=0,\cdots, N$.
Due to the collapse of ensembles \cite{Schillings_2017a,Schillings_2017b,deWiljes2018,chada2019}, the sample variance estimated by ensembles $\{u^{(j)}_N\}_{j=1}^J$ tends to underestimate the true uncertainty. 
EKS \cite{Garbuno-Inigo_2020} generally requires a large number of ensembles to faithfully sample from the true posterior.
Figure \ref{fig:ensbl_std} illustrates that both EKI and EKS with $J=100$ ensembles severely underestimate the posterior standard deviation (plot in the 2d domain) of the parameter function in an elliptic inverse problem (see more details in Section \ref{sec:elliptic}).
Figures \ref{fig:elliptic_std500} and \ref{fig:adif_std500} show more ensembles (500) might improve UQ, but the results highly depend on the specific problem at hand.
Therefore, these methods do not provide a rigorous basis for systematic UQ, especially in high dimensions.
To address this issue, we propose to implement scalable (dimension-independent) inference methods in the sampling step of CES.

\subsection{Sampling -- Infinite-Dimensional MCMC ($\infty$-MCMC)}
\label{sec:geo}
Traditional Metropolis-Hastings algorithms are characterized by deteriorating mixing times upon mesh-refinement or with increasing dimensions.
In contrast, a new class of dimension-independent algorithms -- 
including \emph{preconditioned Crank-Nicolson (pCN)} \cite{cotter13}, \emph{infinite-dimensional MALA ($\infty$-MALA)} \cite{beskos08}, \emph{infinite-dimensional HMC ($\infty$-HMC)} \cite{beskos11}, and \emph{infinite-dimensional manifold MALA ($\infty$-mMALA)} \cite{beskos14} -- has been recently developed. These algorithms are well-defined on the infinite-dimensional Hilbert space,
and thus provide computational benefits with respect to mixing times for finite, but high-dimensional problems in practice.

Consider the following continuous-time Hamiltonian dynamics:
\begin{equation}\label{eq:HD}
\frac{d^2u}{dt^2} + \mK\,
\big\{\, \mC^{-1}u + D\Phi(u) \big\} = 0, \quad \left. \left(v:= \frac{du}{dt}\right)\right|_{t=0} \sim\mN(0,\mK)\ .
\end{equation}
If we let $\mK\equiv\mC$, Equation \eqref{eq:HD} preserves the total energy $H(u,v)=\Phi(u) + \half\Vert v\Vert^2_\mK$.
HMC algorithm \cite{neal10} solves \eqref{eq:HD}
using the following St\"ormer-Verlet symplectic integrator \cite{verlet67}:
\begin{equation}\label{eq:mHDdiscret}
\begin{aligned}
v^- &= v_0 - \tfrac{\alpha\eps}{2}\,\mC D\Phi(u_0)\ ; \\
\begin{bmatrix} u_\eps\\ v^{+}\end{bmatrix} &= \begin{bmatrix} \cos\eps & \sin\eps\\ -\sin\eps & \cos\eps
\end{bmatrix}  \begin{bmatrix} u_0\\ v^{-}\end{bmatrix}\  ;\\
v_\eps &= v^{+} - \tfrac{\alpha\eps}{2}\,\mC D\Phi(u_\eps)\  .
\end{aligned}
\end{equation}
%
Equation \eqref{eq:mHDdiscret} gives rise to the
leapfrog map $\Psi_\eps: (u_{0},v_{0})\mapsto (u_\eps, v_\eps)$.
Given a time horizon $\tau$ and current position 
$u$, the MCMC mechanism proceeds 
by concatenating $I=\lfloor \tau/\epsilon \rfloor$ steps of leapfrog map consecutively:
$u' =\mathcal{P}_u\big\{\Psi_\eps^I(u,v)\big\}, \quad v\sim\mN(0,\mK) $,
where $\mathcal{P}_u$ denotes the projection onto the $u$-argument.
In $\infty$-HMC, the proposal $u'$ is accepted with probability $a(u,u')=1\wedge \exp(-\Delta H(u, v))$
\cite{beskos11}. 
We can use different step-sizes in \eqref{eq:mHDdiscret}: $\eps_1$ for the first and third equations, and $\eps_2$ for the second equation. Setting $I=1$,
$\eps_1^2=h, \cos\eps_2=\frac{1-h/4}{1+h/4}, \sin\eps_2=\frac{\sqrt h}{1+h/4}$, $\infty$-HMC reduces to $\infty$-MALA, which can also be derived from Langevin dynamics \cite{beskos08,beskos2017}.
When $\alpha=0$, $\infty$-MALA further reduces to pCN \cite{beskos2017}, which does not use gradient information and can be viewed as an infinite-dimensional analogue of random walk Metropolis.
%
While the original CES \cite{cleary2020} only uses pCN in the sampling stage, we propose using $\infty$-MALA and $\infty$-HMC with the gradient extracted from NN emulation.

\section{Scaling Up Bayesian UQ with Neural Networks}\label{sec:scaleBUQ}
As mentioned above, there are two major challenges limiting the scalability of Bayesian UQ for inverse problems: 1) intensive computation required for repeated evaluations of likelihood (potential), $\Phi(u)$, and 2) high dimensionality of the discretized space.
If we use $\infty$-MALA or $\infty$-HMC, we also need the gradient $D\Phi(u)$, which is typically not available.
Here, we will use NNs to address these issues. More specifically, we train CNN to emulate the forward evaluation and AE to reduce the parameter dimensionality.
In the following, we discretize the parameter function $u$ into a vector of dimension $d$. 
When there is no confusion, we still denote the discretized parameter vector as $u\in\mbR^d$, which is the input of NNs. We also assume the observation $y\in \mbR^m$.

\subsection{Emulation -- Convolutional Neural Network (CNN)}
\begin{figure}[!t]
\centering
\includegraphics[height=.3\textwidth,width=1\textwidth]{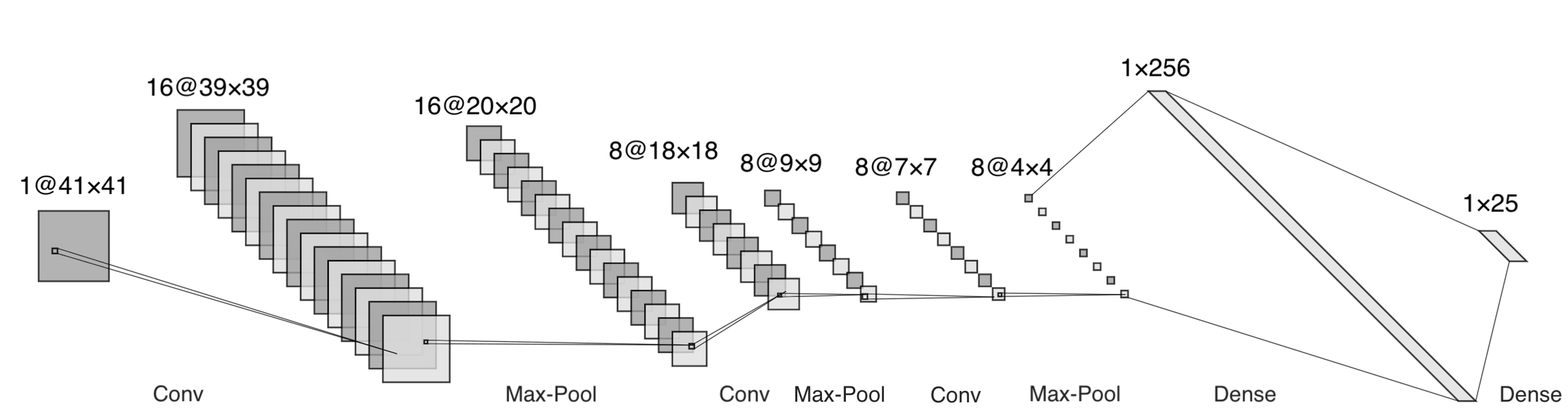}
\vspace{-20pt}
\caption{A typical architecture of convolutional neural network (CNN).}
\label{fig:cnn}
\end{figure}

The ensemble-based algorithms in the calibration phase produce parameters and forward solutions $\{u_n^{(j)}, \mG(u_n^{(j)})\}_{j=1}^J$ for $n=0,\cdots, N$.
These input-output pairs can be used to train a neural network model (DNN or CNN) as an emulator $\mG^e$ of the forward mapping $\mG$ \cite{Goodfellow2016}:
\begin{equation}\label{eq:dnn}
\mG^e(u; \theta) = F_{K-1} \circ \cdots \circ F_0(u), \quad F_k(\cdot) = g_k(W_k \cdot + b_k) \in C(\mbR^{d_k}, \mbR^{d_{k+1}})
\end{equation}
where $d_0=d$, $d_K=m$; $W_k\in \mbR^{d_{k+1}\times d_k}$, $b_k \in \mbR^{d_{k+1}}$, $\theta_k=(W_k, b_k)$, $\theta=(\theta_0,\cdots,\theta_{K-1})$; and $g_k$'s are (continuous) activation functions.
There are multiple choices of activation functions, e.g. $g_k(x)=(\sigma(x_1), \cdots, \sigma(x_{d_{k+1}}))$ with $\sigma\in C(\mbR,\mbR)$ including rectified linear unit (ReLU, $\sigma(x_i)=0\vee x_i$) and 
leaky ReLU ($\sigma(x_i; \alpha)=x_i I(x_i\geq 0)+ \alpha x_i I(x_i< 0)$). 
Alternatively, we can set $g_k(x)=(\sigma_1(x), \cdots, \sigma_{d_{k+1}}(x))\in C(\mbR^{d_{k+1}},\mbR^{d_{k+1}})$, with $\{\sigma_i\}$ defined as softmax: $\sigma_i(x) = e^{x_i}/\sum_{i'=1}^{d_{k+1}} e^{x_{i'}}$.
In our numerical examples, activation functions for DNN/CNN are chosen such that the errors of emulating functions (and their extracted gradients) are minimized.

In many inverse problems, the parameter function $u$ is defined over a 2-d or 3-d field, which possesses unique spatial features resembling an image. This has motivated our choice of CNN for emulators. 
Inspired by biological processes, where the connectivity pattern between neurons resembles the organization of visual cortex \cite{Fukushima_1980},
CNN has become a powerful tool in image recognition and classification \cite{Krizhevsky_2012}.
As a regularized NN with varying depth and width, CNN has much fewer connections and thus fewer training parameters compared to standard fully connected DNN of similar size. Therefore, CNN is preferred to DNN in the CES framework due to its flexibility and computational efficiency.

In general, CNN consists of a series of convolution layers with filters (kernels), pooling layers to extract features, and hidden layers fully connected to the output layer.
The convolutional layer is introduced for sparse interaction, parameter sharing, and equivariant representations \cite{Goodfellow2016}.
At convolutional layer $k$, instead of full matrix multiplication, we use discrete convolution \cite{lecun_1989} with a kernel function $w_k^{(c)}$ defined on one of the $C_k$ feature channels:
\begin{equation}\label{eq:cnn}
F_k(\cdot)=[F_k^{1}(\cdot), \cdots, F_k^{C_k}(\cdot)], \quad F_k^{(c)}(\cdot) = g_k(w_k^{(c)} * \cdot + b_k^{(c)}) \in C(\mbR^{d_k}, \mbR^{d_{k+1}})
\end{equation}
On each channel $c$, the convolution operation $w_k^{(c)}*$ is defined by multiplying its discrete format, a circulant matrix $W_k^{(c)*}$, to its operand \cite{Goodfellow2016,ZHOU_2020}. Together $\{w_k^{(c)}*\}_{c=1}^{C_k}$ amount to a tensor operation.
For each image input, CNN implements the convolution using a sliding window of pre-specified size (kernel size $s\geq 2$) with certain stride (step size) over the image. The resulting operation typically reduces dimension but can be made dimension preserving or expanding through padding \cite{Goodfellow2016}.

After the convolutional layer, a pooling layer (e.g. max-pooling, average-pooling, or sum-pooling) is added to reduce the number of parameters by generating summary statistics (e.g., max, mean, or sum) of the nearby outputs.
Such an operation is a form of non-linear down-sampling that sparsifies the NN but retains the most important information of the input image (function).
Multiple pairs of convolutional and pooling layers (with different configurations) could be concatenated before passing the information to a dense layer to generate forward outputs $\{\mG(u)\}$.
Figure \ref{fig:cnn} illustrates the structure of a CNN used in the elliptic inverse problem (Section \ref{sec:elliptic}).

\begin{figure}[tbp]
  \begin{center}
     \includegraphics[width=.495\textwidth,height=.25\textwidth]{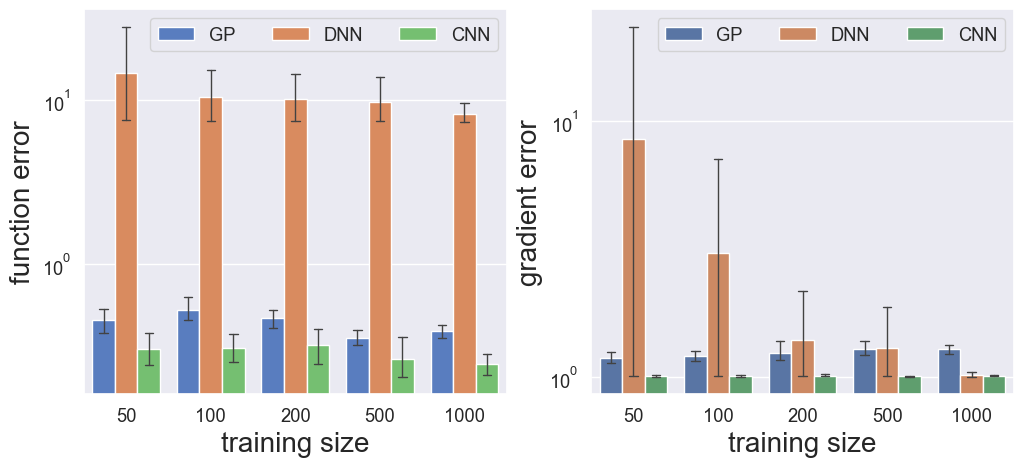}
     \includegraphics[width=.495\textwidth,height=.25\textwidth]{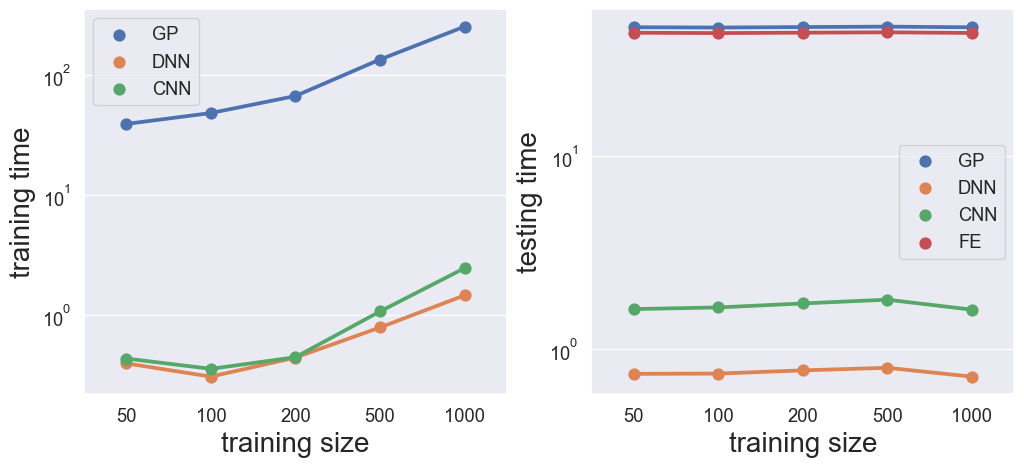}
  \end{center}
  \caption{Comparing the emulation $\mG^e: \mbR^{1681}\to\mbR^{25}$ in an elliptic inverse problem (Section \ref{sec:elliptic}) by GP, DNN and CNN in terms of error (left: function error $\Vert \Phi - \Phi^e\Vert$ and gradient error $\Vert D\Phi - D\Phi^e\Vert$) and time (right).
  Time is also compared with exact calculation of gradients by the finite element method (labeled `FE') using adjoint codes in testing.}
  \label{fig:gp_cnn}
\end{figure}

CES \cite{cleary2020} adopts GP for the emulation step. 
Although infinitely wide NN with Gaussian priors can converge to GP \cite{Neal_1996,Lee_2018,Nalisnick_2018} under certain conditions, the corresponding units do not represent `hidden features' that capture important aspects of the data \cite{Neal_1996}, e.g. the edge of the true log-transmissivity in Figure \ref{fig:truth_obs}, or the truncated Gaussian blob as the true initial condition in Figure \ref{fig:time_soln}.
Here, we show that a finite but potentially deep CNN as an emulator can provide multiple advantages over GP: 1) it is computationally more efficient for large training sets, 2) it is less sensitive to the locations of training samples (even if they are not spread out enough), and 3) it is possible to take advantage of all the ensemble samples collected by EKI or EKS to train CNN without the need to carefully ``design" a training set of controlled size as it is common in GP. 
After the emulator is trained, we could approximate the potential function using the predictions from CNN:
\begin{equation}\label{eq:potential_emu}
\Phi(u^*) \approx \Phi^e(u^*) = \frac{1}{2} \Vert y-\mG^e(u^*)\Vert^{2}_{\Gamma}
\end{equation}
In the sampling stage, the computational complexity could be significantly reduced if we use $\Phi^e$ instead of $\Phi$ in the accept/reject step of MCMC.
If the emulator is a good representation of the forward mapping, then the difference between $\Phi^e$ and $\Phi$ is small. Then, the samples by such emulative MCMC have the stationary distribution with small discrepancy compared to the true posterior $\mu(du)$.

In gradient-based MCMC algorithms, we need to calculate $D\Phi(u)$, i.e., the derivatives of (log) density function $\Phi(u)$ with respective to parameter function $u$.
This is not always available because the forward mapping may not be differentiable (or available) in the solutions of ODE/PDE.
However, (almost everywhere) differentiability is required for each layer of NN as it uses back-propagation \cite{lecun_1989,Hinton_2006a} in the training. Therefore, the gradient of the emulated potential function can be obtained by the chain rule
\begin{equation}\label{eq:dpotential_emu}
D\Phi^e(u^*) = - \langle y-\mG^e(u^*), D\mG^e(u^*)\rangle_{\Gamma}
\end{equation}
where $D\mG^e(u^*)$ can be the output from CNN's back-propagation, e.g. implemented in \texttt{GradientTape} of \texttt{TensorFlow}.
Note that $D\Phi(u^*) \approx D\Phi^e(u^*)$ if $D\Phi(u^*)$ exists.

The universality of deep CNN for continuous functions has been established in \cite{ZHOU_2020}.
This is generalized by the following theorem, which also gives the error bound of CNN emulator in approximating both the true potential $\Phi$ and its gradient $D\Phi$.
To obtain the approximation rate, some regularity conditions are imposed on the target functions being approximated.

\begin{restatable}{thm}{cnnerr}
\label{thm:cnn_err}
Let $2\leq s\leq d$ and $\Omega\subset [-1,1]^d$. Assume $\mG_j\in H^r(\mbR^d) \cap L^\infty([-1,1]^d)$ with $r\geq 1$ such that $v_{\mG_j,2}:=\int_{\mbR^d}\Vert \omega\Vert_1^2|\widehat \mG_j(\omega)| d\omega<\infty$ for $j=1,\cdots, m$.
If $K\geq 2d/(s-1)$, then there exist $\mG^e$ by CNN with ReLU activation function such that
\begin{equation}\label{eq: cnn_errorbound}
\Vert \Phi - \Phi^e\Vert_{H^1(\Omega)} \leq c v_{\mG,2} \sqrt{\log K} K^{-\half - \frac{1}{2d}}
\end{equation}
where we have $\Vert\Phi\Vert_{H^1(\Omega)} = \left(\Vert\Phi\Vert_{L^2(\Omega)}^2+\Vert D\Phi\Vert_{L^2(\Omega)}^2\right)^\half$, $c$ is an absolute constant, and $v_{\mG,2}=\max_{1\leq j\leq m} v_{\mG_j,2}$.
\end{restatable}

\begin{proof}

See Appendix \ref{apx:cnn_err}.
\end{proof}

\begin{rk}
If $r>2+d/2$, then $v_{\mG,2} \leq c\Vert \mG\Vert$. Therefore, we have
\begin{equation}\label{eq: cnn_errorbound1}
\Vert \Phi - \Phi^e\Vert_{H^1(\Omega)} \leq c \Vert \mG\Vert \sqrt{\log K} K^{-\half - \frac{1}{2d}}
\end{equation}
where $\Vert \mG\Vert = \max_{1\leq j\leq m}\Vert \mG_j\Vert_{H^r(\mbR^d)}$ with $\Vert \mG_j\Vert_{H^r(\mbR^d)} := \Vert (1+|\omega|^2)^{r/2}\widehat \mG_j(\omega)\Vert_{L^2(\mbR^d)}$.
\end{rk}

\begin{rk}
By \cite{Barron_1992}, we have a weaker bound with sup-norm under the same condition of Theorem \ref{thm:cnn_err}:
\begin{equation}
\Vert \Phi - \Phi^e\Vert_{W^{1,\infty}(\Omega)} \leq \tilde c \Vert \mG\Vert K^{-\half}
\end{equation}
where $\Vert\Phi\Vert_{W^{1,\infty}(\Omega)} = \max_{0\leq i\leq d} \Vert D_i\Phi\Vert_{L^\infty(\Omega)}$ ($D_0\Phi=\Phi$).
\end{rk}

Even if $D\Phi(u^*)$ does not exist, such gradient information, $D\Phi^e(u^*)$, can still be extracted from the emulator $\mG^e$ to inform the landscape of $\Phi$ in the vicinity of $u^*$.
Note that we train CNN only on $\{u_n^{(j)}, \mG(u_n^{(j)})\}$ as opposed to $\{u_n^{(j)}, D\mG(u_n^{(j)})\}$. That is, no gradient information is used for training.
This is similar to extracting geometric information from GP emulator \cite{stephenson10,lan2016}.
Figure \ref{fig:gp_cnn} compares GP, DNN, and CNN in emulating a forward map that takes a $1681 (41\times 41)$ dimensional discretized parameter $u$ with 25 observations taken from the solution of an elliptic PDE as the output (see more details in Section \ref{sec:elliptic}).
Given limited training data, CNN outperforms both GP and DNN by providing smaller approximation errors ($\Vert \Phi - \Phi^e\Vert$) with lower computational cost.

\subsection{Dimension Reduction -- Autoencoder (AE)}\label{sec:aeDR}
\begin{figure}[!t]
\centering
\includegraphics[height=.3\textwidth,width=1\textwidth]{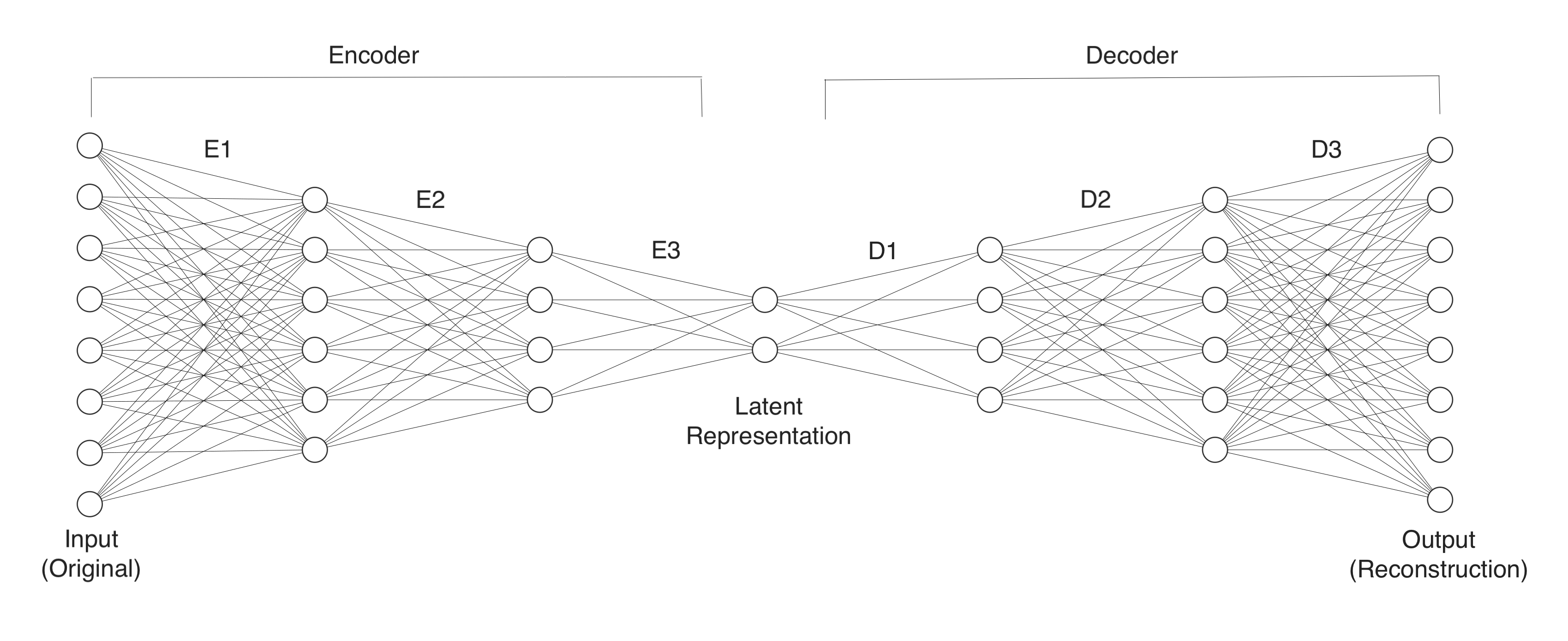}
\vspace{-30pt}
\caption{A typical architecture of autoencoder (AE) neural network.}
\label{fig:ae}
\end{figure}

Although we can reduce computation by emulation, the MCMC algorithms we use for Bayesian inference are still defined in high-dimensional spaces.
In this section, we discuss using AE for dimensionality reduction to further speed up the UQ process \cite{shahbaba2019}.
AE is a special type of feed-forward NN for latent representation learning. The input is encoded into a low-dimensional latent representation (code). The code is then decoded into a reconstruction of the original input (see Figure \ref{fig:ae}). 
The model is trained to minimize the difference between the input and the reconstruction. An AE could learn complicated nonlinear dimensionality reduction. Therefore, it is widely used in many challenging tasks such as image recognition and artificial data generation \cite{Hinton_2006b}. 

While AE is commonly used to reduce the dimensionality of the data, here we use it to reduce the dimensionality of the (discretized) parameter space, still denoted as $\mbX\subset\mbR^d$. Denote the latent space as $\mbX_L$ with dimensionality $d_L\ll d$. Let $u_L\in\mbX_L$ be the latent representation of parameter $u$.
Then the encoder $\phi$ and the decoder $\psi$ are defined respectively as follows
\begin{equation}
\begin{aligned}
\phi: \mbX \to \mbX_L , &\quad u \mapsto u_L \\
\psi: \mbX_L \to \mbX, &\quad u_L \mapsto u_R
\end{aligned}
\end{equation}
where $u_R\in\mbX$ is a reconstruction of $u$; $\phi$ and $\psi$ can be chosen as multilayer NNs similar to Equation \eqref{eq:dnn}.
Depending on the layers and structures, we could have convolutional AE (CAE) \cite{Guo_2017,RIBEIRO_2018}, variational AE (VAE) \cite{Kingma2014,Kingma_2019}, etc. According to universal approximation theorem \cite{Cybenko_1989,Pinkus_1999,Zhou_2017}, a feed-forward artificial NN can approximate any continuous function given some mild assumptions about the activation functions. Theoretically, an AE with suitable activation functions could represent an identity map, i.e. $\psi \circ \phi = id$. An accurate reconstruction of the input implies a good low-dimensional representation encoded in $\phi$.
In practice, the success of the algorithm heavily relies on the quality of the trained AE. 
There is a trade-off in choosing the proper latent dimensionality, $d_L$: smaller $d_L$ throttles the information flowing through AE and leads to higher reconstruction errors (See Figure \ref{fig:bbd_pairpdf}); larger $d_L$ could reduce reconstruction error, but it may also negatively impact the computational efficiency of MCMC algorithms defined on the resulting latent subspace.
Note that we train the AE with ensembles $\{u_n^{(j)}\}_{j=1,n=0}^{J,N}$ from the calibration stage.
Even though $\psi\circ\phi$ might be different from the identity map $id$, AE could still provide a reconstruction $\psi\circ\phi(u)$ very close to the original parameter $u$. See Figure \ref{fig:elliptic_reconstruction} (Section \ref{sec:elliptic}) and Figure \ref{fig:adif_reconstruction} (Section \ref{sec:adif}) for examples.

The potential function $\Phi(u)$ and its derivative $D\Phi(u)$ can be projected to the latent space $\mbX_L$ -- denoted as $\Phi_r(u_L)$ and $D\Phi_r(u_L)$ respectively -- as follows:
\begin{equation}\label{eq:potential_lat}
\begin{aligned}
\Phi_r(u_L) &= \Phi(u) = \Phi(\psi(u_L)) \\
D\Phi_r(u_L) &= \left(\frac{\pa u}{\pa u_L}\right)^T \frac{\pa \Phi(u)}{\pa u} = (d\psi(u_L))^T D\Phi(\psi(u_L))
\end{aligned}
\end{equation}
where $d\psi=\frac{\pa u}{\pa u_L}$ is the Jacobian matrix of size $d\times d_L$ for the decoder $\psi$. 
The derivative information $D\Phi_r(u_L)$ needed in the gradient-based MCMC ($\infty$-MALA and $\infty$-HMC) will be discussed in Section \ref{sec:dream}.
In practice, we avoid explicit computation of the Jacobian matrix $d\psi$ by calculating the Jacobian-vector action altogether:
$D\Phi_r(u_L) = \left. \frac{\pa}{\pa u} \right|_{u=u_L} [\psi(u)^T D\Phi(\psi(u_L)) ]$,
which is obtained from AE's back-propagation.


Now we are ready to combine emulation with dimension reduction to further improve the computation efficiency. 
The resulting approximate MCMC algorithms in the latent space involve potential function and its derivative, denoted as $\Phi^e_r(u_L)$ and $D\Phi^e_r(u_L)$ respectively,
which are defined by replacing $\Phi$ with $\Phi^e$ in Equation \eqref{eq:potential_lat}.

\subsection{Dimension Reduced Emulative Autoencoder MCMC (DREAMC)}\label{sec:dream}
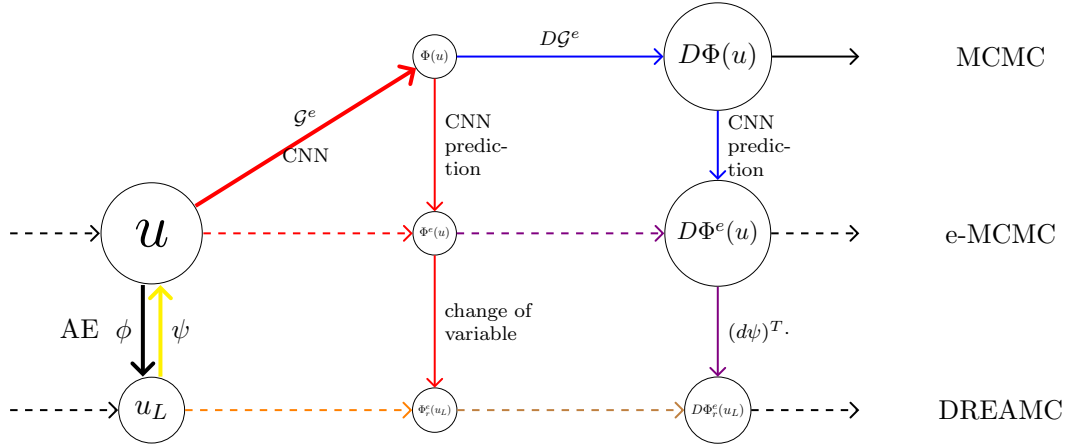
\begin{figure}[t]
\centering
\begin{tikzpicture}
\node[shape=circle,draw,scale=1.1] (uL) at (0,0) {$u_L$};
\node[shape=circle,draw,scale=.4] (phier) at (4,0) {$\Phi^e_r(u_L)$};
\node[shape=circle,draw,scale=.5] (dphier) at (8,0) {$D\Phi^e_r(u_L)$};
\node[draw=none] at (12,0) {DREAMC};
\path[-angle 90,thick]
(-2,0) edge [dashed] (uL)
(uL) edge [dashed,draw=orange] (phier)
(phier) edge [dashed,draw=brown] (dphier)
(dphier) edge [dashed] (10,0);

\node[shape=circle,draw,scale=2.25] (u) at (0,2.5) {$u$};
\node[shape=circle,draw,scale=.45] (phie) at (4,2.5) {$\Phi^e(u)$};
\node[shape=circle,draw,scale=.9] (dphie) at (8,2.5) {$D\Phi^e(u)$};
\node[draw=none] at (12,2.5) {e-MCMC};
\path[-angle 90,thick]
(-2,2.5) edge [dashed] (u)
(u) edge [dashed,draw=red] (phie)
(phie) edge [dashed,draw=violet] (dphie)
(dphie) edge [dashed] (10,2.5);
\draw[-angle 90,ultra thick,shift right=.75ex] 
(u) edge node[left] {AE \;$\phi$} (uL)
(uL) edge[draw=yellow] node[right] {$\psi$} (u);
\path[-angle 90,thick,font=\scriptsize] (phie) edge[draw=red] node[right,text width=.5in] {change of variable} (phier);
\path[-angle 90,thick,font=\scriptsize] (dphie) edge[draw=violet] node[right] {$(d\psi)^T\cdot$} (dphier);

\node[shape=circle,draw,scale=.5] (phi) at (4,5) {$\Phi(u)$};
\node[shape=circle,draw] (dphi) at (8,5) {$D\Phi(u)$};
\node[draw=none] at (12,5) {MCMC};
\path[-angle 90,ultra thick,font=\scriptsize] (u) edge[draw=red] node[above] {$\mG^e$} node[below]{CNN} (phi);
\path[-angle 90,thick,font=\scriptsize]
(phi) edge[draw=blue] node[above] {$D\mG^e$} (dphi)
(dphi) edge (10,5)
(phi) edge[draw=red] node[right,text width=.5in] {CNN prediction} (phie)
(dphi) edge[draw=blue] node[right,text width=.5in] {CNN prediction} (dphie);
\end{tikzpicture}
\caption{\small Relationship among quantities in various MCMC algorithms. Node sizes indicate relative dimensions of these quantities. Thick solid arrows mean training neural networks.
              Dashed arrows with colors represent mappings that are not directly calculated but actually have equivalent compositions indicated by the same color, e.g. $u\mapsto \Phi^e(u)$ (dashed red arrow) obtained by training CNN (thick solid red arrow) followed by network prediction (solid red arrow); or by color mixing, e.g. $u_L\mapsto \Phi^e_r(u_L)$ (dashed orange arrow) as a result of combing the decoder $\psi$ (thick solid yellow arrow), $u\mapsto \Phi^e(u)$ (dashed red arrow), and the change of variable (solid red arrow).}
\label{fig:algs}
\end{figure}

Next, we combine all the techniques discussed above to speed up Bayesian UQ for inverse problems.
More specifically, 
our main method is composed of the following three stages:
\begin{enumerate}[itemsep=0pt]
	\item {\bf Calibration}: collect $JN$ samples $\{u_n^{(j)}, \mG(u_n^{(j)})\}_{j,n}$ from EKI or EKS procedure;
	\item {\bf Emulation}: build an emulator of the forward mapping $\mG^e$ based on $\{u_n^{(j)}, \mG(u_n^{(j)})\}_{j,n}$ (and extract $D\mG^e$) using CNN; train an AE $(\phi,\psi)$ based on $\{u_n^{(j)}\}_{j,n}$;
	\item {\bf Sampling}: run approximate MCMC based on emulation to propose $u'$ from $u$:
	\begin{enumerate}[noitemsep,label=\roman*)]
		\item obtain the projection of $u$ by $u_L = \phi(u)$;
		\item propose $u'_L$ from $u_L$ by $\infty$-MCMC (with $\Phi^e_r$ and $D\Phi^e_r$) in the latent space $\mbX_L$;
		\item obtain the sample $u' = \psi(u'_L)$
	\end{enumerate}
\end{enumerate}
Within the class of $\infty$-MCMC, we can use the emulated potential and its derivative instead of exact calculation. We refer to the resulting algorithms as \emph{emulative $\infty$-MCMC (e-MCMC)}.
Further, we can use AE to project these approximate MCMC into low-dimensional latent space. We denote these algorithms as \emph{dimension-reduced emulative autoencoder $\infty$-MCMC (DREAMC)}.
Figure \ref{fig:algs} illustrates the relationship among various quantities involved in these MCMC algorithms.

We note that if we accept/reject proposals $u'_L$ in the latent space with $\Phi^e_r$, then there is no need to traverse between the original space and the latent space constantly.
The chain can mainly stay in the latent space $\mbX_L$ to collect samples $\{u_L\}$, as shown in the bottom level of Figure \ref{fig:algs}, and move back to the original space $\mbX$ when relevant emulated quantities are needed.
In the following, we describe the details of DREAMC algorithms.

For the convenience of following disposition, we first whiten the coordinates by the transformation $u \mapsto \tilde u:=\mC^{-\half} u$.
The whitened variable $\tilde u$ has the prior $\tilde \mu_0 = \mathcal N(0, \mI)$, where the identity covariance operator is not a trace-class on $\mbX$.
However, random draws from $\tilde\mu_0$ are square-integrable in the weighted space $\mathrm{Im}(\mC^{-\half})$.
We can still obtain a well-defined function space proposal for parameter $u$ after inverting the transformation \cite{cui16,LAN2019a}.
In the whitened coordinates $\tilde u$, the Langevin and 
Hamiltonian \eqref{eq:HD} dynamics (with algorithmic parameter $\alpha\equiv 1$) can be written as follows respectively:
\begin{align}
& \frac{d\tilde u}{dt} = -\half\,\big\{ \mI \tilde u+ \alpha D\Phi(\tilde u)\big\} + \frac{dW}{dt} \label{eq:LD_white}\\
& \frac{d^2\tilde u}{dt^2} + \big\{\, \mI \tilde u + \alpha D\Phi(\tilde u) \big\} = 0, \quad \left. \left(\tilde v:= \frac{d\tilde u}{dt}\right)\right|_{t=0} \sim\mN(0,\mI)\ . \label{eq:HD_white} 
\end{align}
where $D\Phi(\tilde u)=\mC^\half D\Phi(u)$.
Then, we can train CNN using $\{\tilde u^{(j)}_n, \mG(\tilde u^{(j)}_n)\}_{j,n}$ and AE using $\{\tilde u^{(j)}_n\}_{j,n}$.

On the other hand, since the AE model does not preserve the volume ($\psi\circ\phi\approx id$ but $\psi\circ\phi\neq id$), the acceptance of proposals in the latent space needs to be adjusted with a volume correction term $\frac{V'}{V}$ in order to maintain the ergodicity \cite{lan2014,shahbaba2019}.
Note, the volume adjustment term $\frac{V'}{V}$ breaks into the product of Jacobian determinants of the encoder $\phi$ and the decoder $\psi$ that can be calculated with Gramian function as follows \cite{shahbaba2019}:
\begin{equation}\label{eq:volx}
\frac{V'}{V} = \det (d\psi(\tilde u'_L)) \det (d\phi(\tilde u)) = \sqrt{\det \left[ \left(\frac{\pa \tilde u'}{\pa \tilde u'_L}\right)^T \left(\frac{\pa \tilde u'}{\pa \tilde u'_L}\right) \right]} \sqrt{\det \left[ \left(\frac{\pa \tilde u_L}{\pa \tilde u}\right) \left(\frac{\pa \tilde u_L}{\pa \tilde u}\right)^T \right]}
\end{equation}
where terms under square root are determinants of matrices with small size $d_L\times d_L$, which can be obtained by the singular value decomposition of the Jacobian matrices respectively.
In practice, we can exclude $\frac{V'}{V}$ from the acceptance probability and use it as a resampling weight as in importance sampling \cite{lan2014a}. Alternatively, we can ignore the accept/reject step for an approximate Bayesian UQ \cite{welling11,shahbaba2019}.

To derive $\infty$-HMC in the latent space based on the Hamiltonian dynamics in the whitened coordinates \eqref{eq:HD_white}, we also need to project $\tilde v\sim\mN(0,I_d)$ into $d_L$-dimensional latent space $\mbX_L$.
We could use the same encoder $\phi$ as in \cite{shahbaba2019}; however, since $\tilde v_i\overset{iid}{\sim} \mN(0,1)$ for $i=1,\cdots, d$, we just set $\tilde v_L\sim\mN(0,I_{d_L})$ for simplicity.
Then, the $\infty$-HMC proposal $\Psi_\eps: (\tilde u_{L,0},\tilde v_{L,0})\mapsto (\tilde u_{L,\eps}, \tilde v_{L,\eps})$ in the whitened augmented latent space with emulated gradient becomes
\begin{equation}\label{eq:mHDdiscret_white}
\begin{aligned}
\tilde v^-_L &= \tilde v_{L,0} - \tfrac{\alpha\eps}{2}\, D\Phi^e_r(\tilde u_{L,0})\ ; \\
\begin{bmatrix} \tilde u_{L,\eps}\\ \tilde v^{+}_L\end{bmatrix} &= \begin{bmatrix} \cos\eps & \sin\eps\\ -\sin\eps & \cos\eps
\end{bmatrix}  \begin{bmatrix} \tilde u_{L,0}\\ \tilde v^{-}_L\end{bmatrix}\  ;\\
\tilde v_{L,\eps} &= \tilde v^{+}_L - \tfrac{\alpha\eps}{2}\, D\Phi^e_r(\tilde u_{L,\eps})\  .
\end{aligned}
\end{equation}
The acceptance probability for the resulting DREAMC-$\infty$-HMC algorithm involves $H(\tilde u_L,\tilde v_L)=\Phi^e_r(\tilde u_L) + \half\Vert \tilde v_L\Vert^2$ and becomes $a(\tilde u_L,\tilde u'_L)=1\wedge \exp(-\Delta H(\tilde u_L, \tilde v_L))\frac{V'}{V}$ with $\frac{V'}{V}$ as in \eqref{eq:volx} and
\begin{equation}\label{eq:acpt_dreHMC}
\begin{aligned}
\Delta H(\tilde u_L, \tilde v_L) =& H(\Psi_\eps^I(\tilde u_L,\tilde v_L)) - H(\tilde u_L,\tilde v_L) \\
=& \Phi(\tilde u_{L,I}) - \Phi(\tilde u_{L,0}) - \frac{\alpha^2\eps^2}{8}\left\{ \Vert D\Phi^e_r(\tilde u_{L,I})\Vert^2 - \Vert D\Phi^e_r(\tilde u_{L,0})\Vert^2 \right\} \\
 &- \frac{\alpha\eps}{2} \sum_{i=0}^{I-1} (\langle \tilde v_{L,i}, D\Phi^e_r(\tilde u_{L,i}) \rangle + \langle \tilde v_{L,i+1}, D\Phi^e_r(\tilde u_{L,i+1}) \rangle)
\end{aligned}
\end{equation}
We summarize DREAMC-$\infty$-HMC in Algorithm \ref{alg:DRe-infHMC} in Appendix \ref{apx:alg}, which includes DREAMC-$\infty$-MALA with $I=1$ and DREAMC-pCN with $\alpha=0$ as special cases.

\section{Illustrations}\label{sec:illust}
In this section, we investigate two low-dimensional inverse problems: a three-dimensional linear Gaussian inverse problem with tractable posterior distribution and a four-dimensional nonlinear banana-biscuit-doughnut (BBD) distribution \cite{lan2016} with complex geometry.
We illustrate the validity of our proposed emulative $\infty$-MCMC and DREAMC algorithms using these two examples.
The structure of our NN models (e.g., the number of layers and choice of activation functions) in this section and next section are chosen from a small subset of options to minimize the overall error; therefore, they may not optimal globally.
All computer codes are available at \texttt{GitHub} \href{https://github.com/lanzithinking/DREAMC-BUQ}{\underline{https://github.com/lanzithinking/DREAMC-BUQ}}.

\begin{figure}[t]
	\centering
	\includegraphics[height=.45\textwidth,width=1\textwidth]{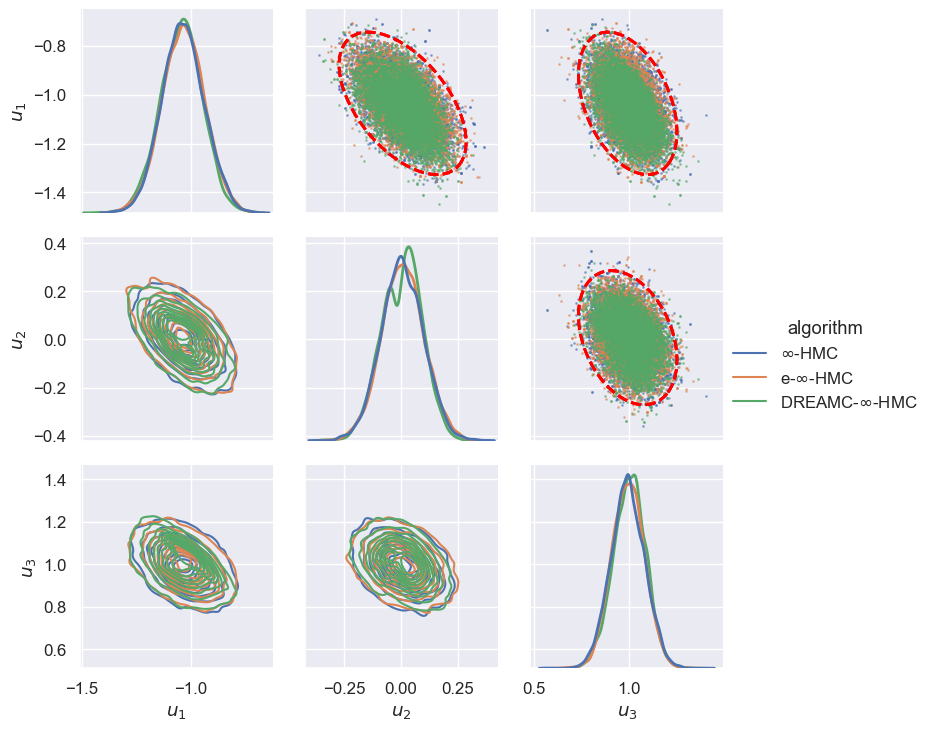}
	\vspace{-10pt}
	\caption{Linear Gaussian inverse problem: pairwise marginal posterior density estimation.}
	\label{fig:lin_pairpdf}
\end{figure}
\subsection{Linear Gaussian inverse problem}

We first consider the following linear Gaussian inverse problem.
\begin{align*}
y &=  \mG(u)+ \eta, \quad \mG(u) = A u, \quad \eta \sim \mN (0, \Gamma) \\
u &\sim \mN(0, \Sigma_0)
\end{align*} 
where $A$ is a matrix of size $m\times d$.
In this example, we set $\Gamma= \sigma^2_\eta I_m$ with $\sigma^2_\eta=0.1$ and $\Sigma_0=\sigma^2_u I_d$ with $\sigma^2_u=1$.
We randomly generate $A$ with $a_{ij}\sim \mathrm{unif}[0,1]$. Further, we assume $d=3$ and set the true value $u^\dagger=(-1,0,1)$.
We generate $m=100$ data points, $y=\{y_n\}_{n=1}^m$, with $\mG(u^\dagger)=A u^\dagger$.
The inverse problem involves finding the posterior distribution of $u|y$ which has the following analytic form:
\begin{align*}
u | y &\sim \mN(\mu, \Sigma),  \quad 
\mu = \Sigma \tp A \Gamma^{-1} y , \quad \Sigma^{-1}=\Sigma_0^{-1} + \tp A \Gamma^{-1} A
\end{align*}

We follow the procedure of CES outlined in Section \ref{sec:dream}. First, we run EKS with ensemble size $J=100$ for $N=50$ iterations and collect 5000 ensemble pairs $\{u_n^{(j)}, \mG(u_n^{(j)})\}_{j=1,n=1}^{J,N}$.
Then we train DNN with $75\%$ of these ensembles and use the remaining $25\%$ for testing. The DNN has 3 layers with `softplus' activation function for the hidden layers and `linear' activation for the output layer with the units linearly interpolated between input dimension ($d=3$) and output dimension ($m=100$).
We also train AE with the same split of training/testing data. The AE has latent dimension $d_L=2$, with 2 encoder layers and 2 decoder layers. We use `LeakyReLU($\alpha=2$)' as the activation function.
We run $\infty$-HMC, emulative $\infty$-HMC and DREAMC $\infty$-HMC (Algorithm \ref{alg:DRe-infHMC}) to collect 10000 posterior samples of $u$ after burning in the first 10000 respectively.
Figure \ref{fig:lin_pairpdf} shows that both emulative $\infty$-HMC and DREAMC $\infty$-HMC generate samples very close to the original $\infty$-HMC. All three methods capture the true distribution whose 3-standard deviation contours are plotted as red ellipses.

\subsection{Nonlinear Banana-Biscuit-Doughnut (BBD) inverse problem}

\begin{figure}[t] 
   \centering
   \includegraphics[width=1\textwidth,height=.3\textwidth]{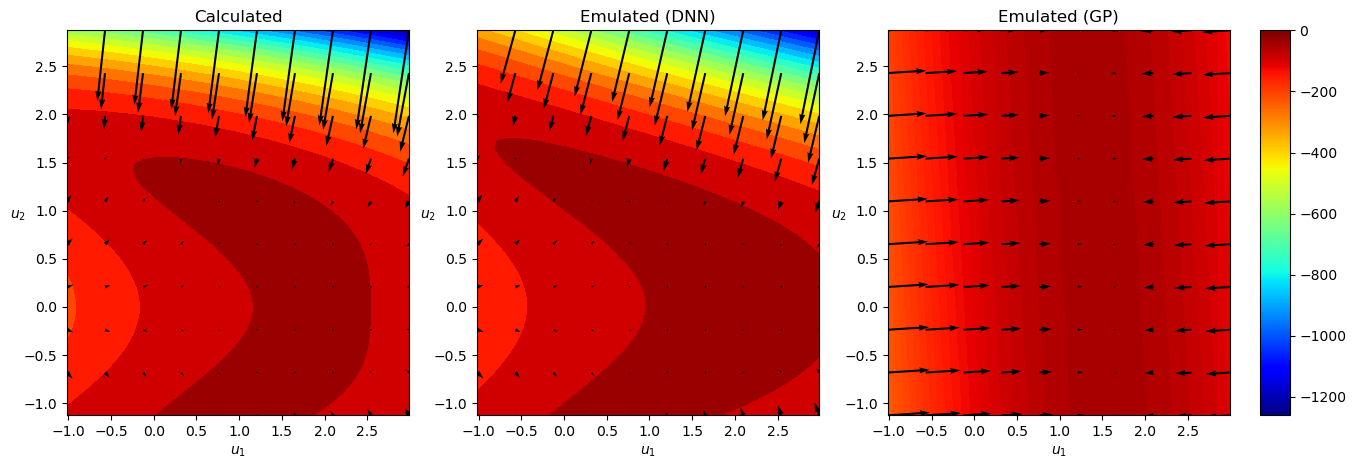} 
   \caption{Nonlinear BBD inverse problem: comparing the true (left) potential (colored contours) and its gradient (arrows) with the corresponding emulated quantities given by DNN (middle) and GP (right).}
   \label{fig:bbd_emulation}
\end{figure}

Next, we challenge our proposed methodology with a complex four-dimensional Banana-Biscuit-Doughnut (BBD) distribution \cite{lan2016}.
BBD distribution was first proposed in \cite{lan2016} as a benchmark for testing MCMC algorithms and has recently been revisited by \cite{shahbaba2019}.
The name of the distribution comes from 3 possible shapes of pairwise marginal distributions resembling a banana in (1,2) dimension, a biscuit in (1,3) dimension, and a doughnut in (2,4) dimension (see Figure \ref{fig:bbd} in Appendix \ref{apx:ext}).
It can be cast into a Bayesian inverse problem with parameters $u=(u_1,\cdots,u_d)$:
\begin{align*}
y &=  \mG(u)+ \eta, \quad \eta \sim \mN (0, \sigma^2_\eta I_m) \\
\mG(u) &= A \mathcal S u, \quad  \mathcal S u = \left(u_1, u_2^2, \cdots, u_k^{p_k}, \cdots,u_d^{p_d}\right), \quad p_k= 2 -(k \; \mathrm{mod} \; 2) \\
u &\sim \mN(0, \sigma^2_u I_d)
\end{align*} 
where $A$ is a matrix of size $m\times d$. Therefore, $\mG: \mathbb R^d \to \mathbb R^m$ is a linear forward mapping of $\mathcal S u$ but a non-linear operator of $u$.

As before, we generate a matrix $A$ and true vector $u^\dagger$ with elements being random integers between 0 and $d$. Then, we obtain $m=100$ data points, $\{y_n\}_{n=1}^m$, with $\mG(u^\dagger)=A \mathcal S u^\dagger, \sigma^2_\eta=1$. In the prior, we set $\sigma^2_u=1$.
The inverse problem involves finding $u$ for given data $\{y_n\}_{n=1}^m$.
The distribution has a complex geometric structure, which makes it challenging for MCMC algorithms to explore (Figure \ref{fig:bbd} in Appendix \ref{apx:ext}). 

We collect training samples, $\{u_n^{(j)}, \mG(u_n^{(j)})\}_{j=1,n=1}^{J,N}$, using EKS with $J=100$ ensembles for $N=50$ iterations.
Then, we train DNN (with a similar structure as before, but with 5 layers) and AE (with the same configuration as before, but with latent dimension $d_L=3$).
For comparison, we also train GP with anisotropic squared exponential kernel based on these 5000 ensembles. The GP model is coded in \texttt{GPflow}, which uses scalable variational GP (SVGP) algorithms \cite{Opper_2009, Hensman_2015}. SVGP determines the value of hyper-parameters (e.g. length scales) by maximizing the evidence lower bound (ELBO). In \texttt{GPflow}, stochastic optimizers such as stochastic gradient descent (SGD) and adaptive moment estimation (ADAM) can be directly used. The set-up for choosing mini-batches in our GP training is the same as our DNN training.
Both emulators, $\mG^e: \mathbb R^4\to \mathbb R^{100}$, are built for $A \mathcal S$, and the log-likelihood is computed using \eqref{eq:potential_emu}.
Figure \ref{fig:bbd_emulation} shows that the emulation by DNN (middle) is much closer to the truth (left) than that of GP (right).
As we can see from these two illustrative examples, GP works equally well as DNN in emulating the low-dimensional linear map, but it performs much worse in the high-dimensional (many-output) nonlinear problems.

\begin{figure}[t]
	\centering
	\includegraphics[height=.45\textwidth,width=1\textwidth]{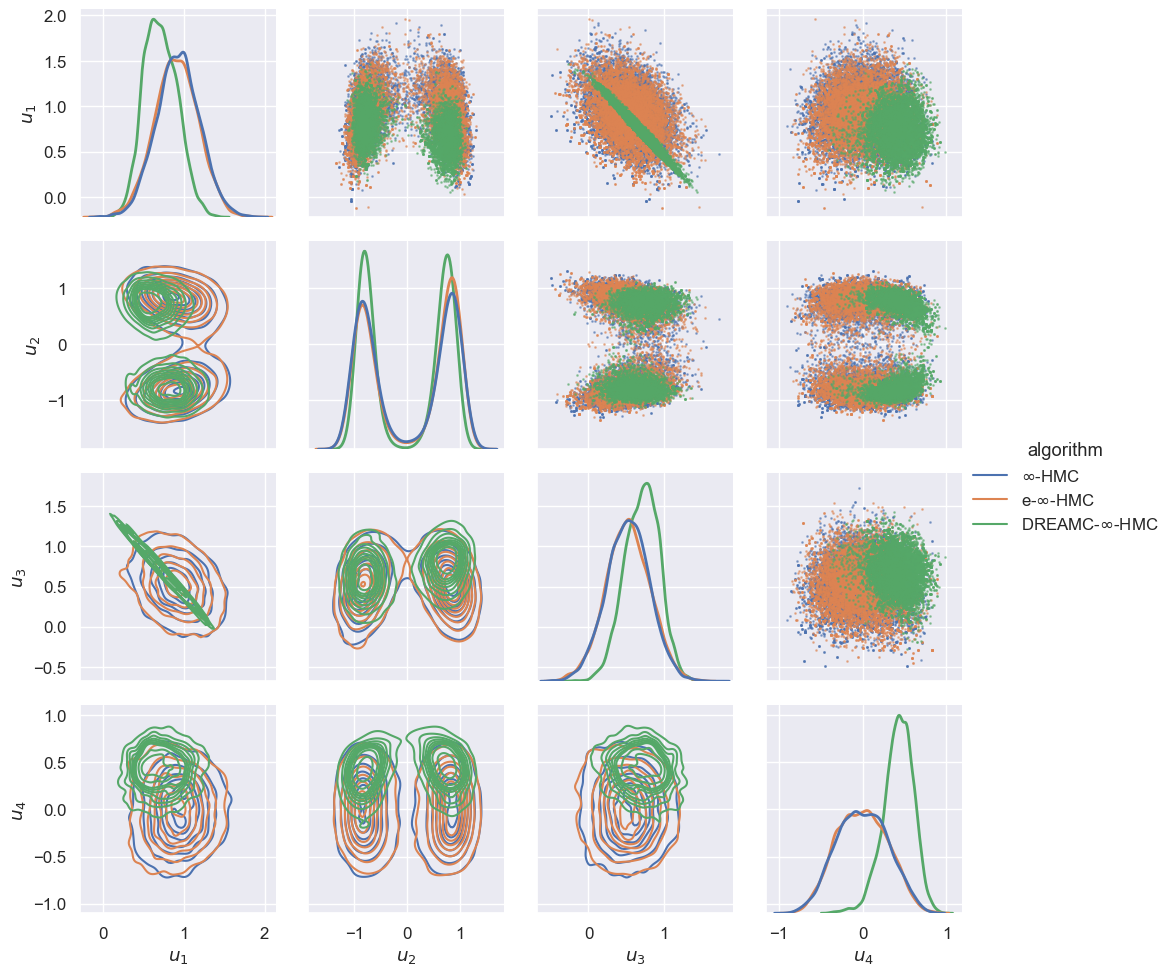}
	\vspace{-10pt}
	\caption{BBD inverse problem: pairwise marginal posterior density estimation.}
	\label{fig:bbd_pairpdf}
\end{figure}
Next, we run $\infty$-HMC, emulative $\infty$-HMC and DREAMC $\infty$-HMC to collect 10000 posterior samples of $u$ after discarding the first 10000.
We plot the marginal distributions of model parameters estimated by these MCMC samples in Figure \ref{fig:bbd_pairpdf}.
The resulting distribution from emulative $\infty$-HMC is close to that of $\infty$-HMC.
In this example, the intrinsic dimension (the dimension of space containing essential data information) is $d=4$ but the latent dimension is $d_L=3$.
Therefore, some information is lost in AE. This leads to the discrepancy between DREAMC $\infty$-HMC and the other two algorithms.
Still, the DREAMC algorithm recovers enough details of the posterior. 

\section{Numerical Experiments}\label{sec:numerics}

In this section, we consider two high-dimensional inverse problems involving elliptic PDE and advection-diffusion equation.
In both problems, the forward parameter-to-observation mappings are nonlinear, and the posterior distributions are non-Gaussian.
The high dimensionality of the discretized parameter imposes a big challenge on Bayesian UQ. The second inverse problem involving advection-diffusion equation is even more challenging because it is based on spatiotemporal observations. We demonstrate substaintial numerical advantages of our proposed methods and show that they indeed can scale up the Bayesian UQ for PDE-constrained inverse problems to thousands of dimensions.

\subsection{Elliptic Inverse Problem}\label{sec:elliptic}

\begin{figure}[t]
\centering
\includegraphics[width=1\textwidth,height=.3\textwidth]{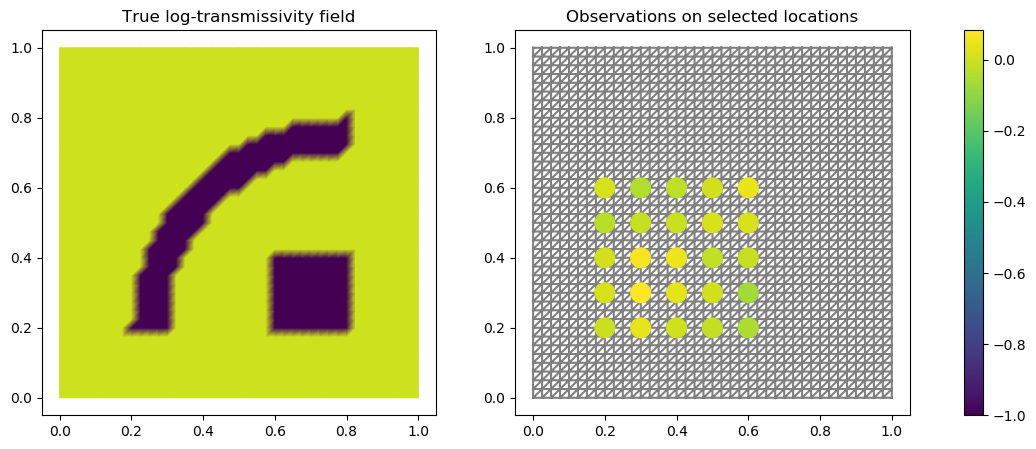}
\caption{Elliptic inverse problem: true log-transmissivity field $u^\dagger(\bx)$ (left), and $25$ observations on selected locations indicated by circles (right), with color indicating their values.}
\label{fig:truth_obs}
\end{figure}

The following elliptic PDE \cite{cui16,LAN2019a} is defined on the unit square domain $\Omega=[0,1]^2$:
\begin{equation}\label{eq:elliptic}
\begin{aligned}
-\nabla \cdot (k(\bx) \nabla p(\bx)) &= f(\bx), \quad \bx \in \Omega \\
\langle k(\bx) \nabla p(\bx), \vec n(\bx) \rangle &= 0, \quad \bx \in \pa\Omega \\
\int_{\pa\Omega} p(\bx) dl(\bx) &= 0
\end{aligned}
\end{equation}
where $k(\bx)$ is the transmissivity field, $p(\bx)$ is the potential function, $f(\bx)$ is the forcing term,
and $\vec n(\bx)$ is the outward normal to the boundary.
The source/sink term $f(\bx)$ is defined by the superposition of four weighted Gaussian plumes with standard deviation $0.05$, 
centered at $\bx = [0.3, 0.3],\, [0.7, 0.3],\, [0.7, 0.7],\, [0.3, 0.7]$, with weights $\{2, -3, 3, -2\}$ respectively, as shown in the left panel of Figure \ref{fig:force_soln}.

The transmissivity field is endowed with a log-Gaussian prior, i.e.
\begin{equation*}
k(\bx) = \exp(u(\bx)), \quad u(\bx) \sim \mathcal N(0, \mC)
\end{equation*}
where the covariance operator $\mC$ is defined through an exponential kernel function
\begin{equation*}
\mC: \mbX \rightarrow \mbX, \; u(\bx) \mapsto \int c(\bx, \bx') u(\bx') d\bx', \quad c(\bx, \bx') = \sigma_u^2 \exp \left( -\frac{\Vert \bx- \bx' \Vert}{2\ell}\right), \,\textrm{for}\; \bx,\bx' \in \Omega
\end{equation*}
with the prior standard deviation $\sigma_u=1.25$ and the correlation length $\ell=0.0625$.
To make the inverse problem more challenging, we follow \cite{cui16} to use a true log transmissivity field $u^\dagger(\bx)$ that is not drawn from the prior,
as shown in the left panel of Figure \ref{fig:truth_obs}.
The right panel of Figure \ref{fig:force_soln} shows the potential function, $p(\bx)$, solved
with $u^\dagger(\bx)$, which is also used for generating noisy observations.
Partial observations are obtained by solving $p(\bx)$ on an $81\times 81$ mesh and then collecting at $25$ measurement sensors $\{\bx_i\}_{i=1}^{25}$ as shown by the circles on the right panel of Figure \ref{fig:truth_obs}. 
The corresponding observation operator $\mathcal O$ yields the data $y\in\mbR^{25}$
\begin{equation*}
y = \mathcal O p(\bx) + \eta, \quad \eta \sim \mathcal N(0, \sigma_\eta^2 I_{25})
\end{equation*}
where we consider the signal-to-noise ratio $\textrm{SNR}=\max_\bx \{u(\bx)\}/\sigma_\eta=50$ in this example.

\begin{figure}[t]
\begin{subfigure}[b]{1\textwidth}
\includegraphics[width=1\textwidth,height=.25\textwidth]{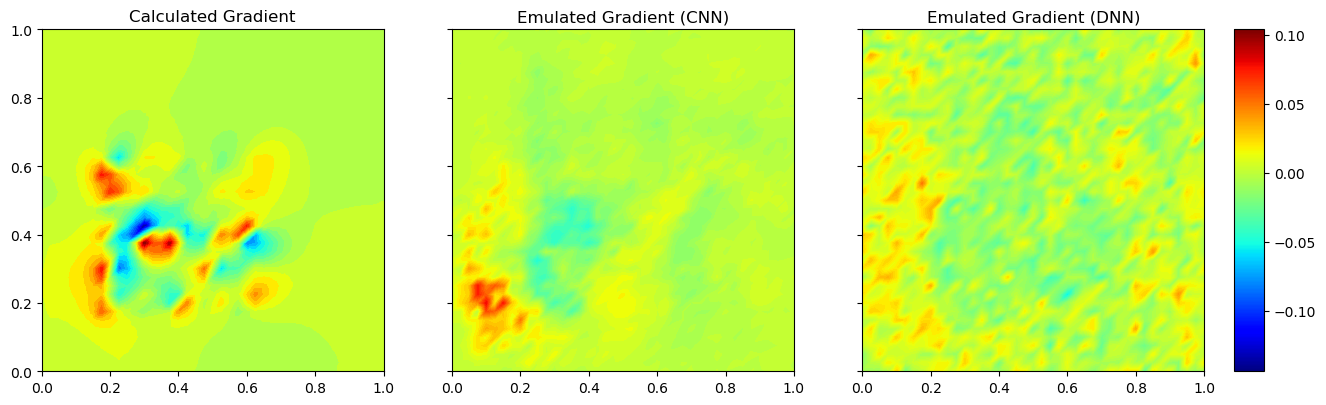}
\caption{CNN (middle) and DNN (right) emulation ($\mG^e: \mbR^{1681}\to\mbR^{25}$) extracting gradients $D\Phi^e(u^\text{MAP})$ compared with the true gradient $D\Phi(u^\text{MAP})$ (left).}
\label{fig:elliptic_extrctgrad}
\end{subfigure}
\begin{subfigure}[b]{1\textwidth}
\includegraphics[width=1\textwidth,height=.25\textwidth]{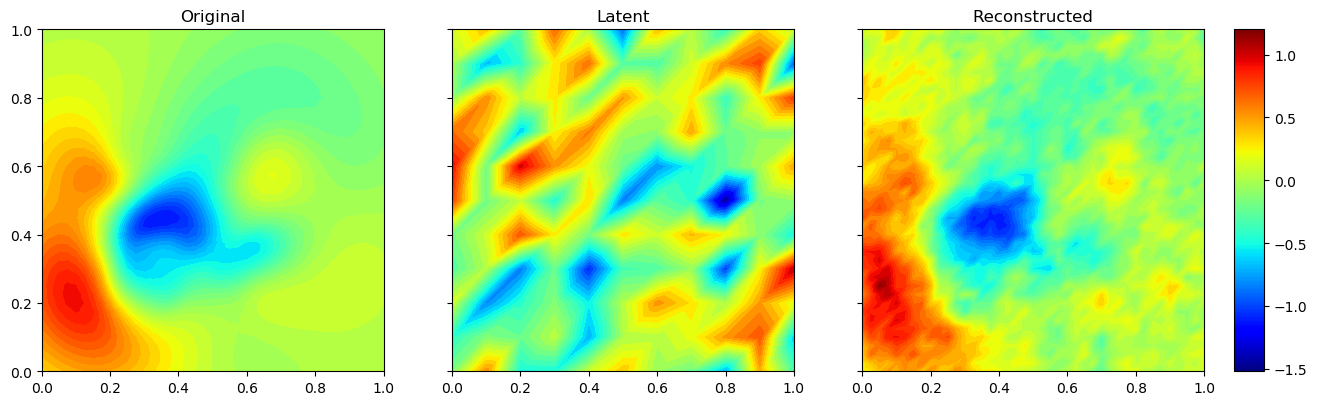}
\caption{AE compressing the original function $u^\text{MAP}$ (left) into latent space $u^\text{MAP}_r$ (middle) and reconstructing it in the original space $u^{\text{MAP}'}$ (right).}
\label{fig:elliptic_reconstruction}
\end{subfigure}
\vspace{-20pt}
\caption{Elliptic inverse problem: outputs by NNs viewed as 2d images.}
\end{figure}

The inverse problem involves sampling from the posterior of the log-transmissivity field $u(\bx)$, which becomes a vector with dimension of $1681$ after being discretized on $41\times 41$ mesh (with Lagrange degree $1$). We use the CES framework described in Section \ref{sec:dream}. 
In the calibration stage, we collect $\{u_n^{(j)}, \mG(u_n^{(j)})\}_{j=1,n=1}^{J,N}$ from $N=10$ iterations of EKS runs with ensemble size $J=500$.
For the emulation, we train DNN and CNN with $75\%$ of these 5000 ensembles and test/validate them with the remaining $25\%$.
The DNN has 3 layers with `softplus'  activation function for the hidden layers and `linear' activation for the output layer and $40\%$ nodes dropped out.
The structure of CNN is illustrated in Figure \ref{fig:cnn} with `softplus' activation for the convolution layers, `softmax' activation for the latent layer (dimension 256) and `linear' activation for the output layer. The trained CNN has drop out rate of $50\%$ on all its nodes.
Figure \ref{fig:elliptic_extrctgrad} compares the true gradient function $D\Phi(u^\text{MAP})$ (left panel) and its emulations $D\Phi^e(u^\text{MAP})$ (middle and right panels) as in Equation \eqref{eq:dpotential_emu} extracted from two types of NNs. These gradient functions are plotted on the 2d domain $[0,1]^2$. We can see that even trained on forward outputs without any gradient information, these extracted gradients provide decent approximations to the true gradient capturing its main graphical feature viewed as a 2d image. The result by CNN is qualitatively better than DNN, which is supported by the numeric evidence of error comparison illustrated in the left panel of Figure \ref{fig:gp_cnn}.

\begin{figure}[t]
\begin{subfigure}[b]{1\textwidth}
\includegraphics[width=1\textwidth,height=.46\textwidth]{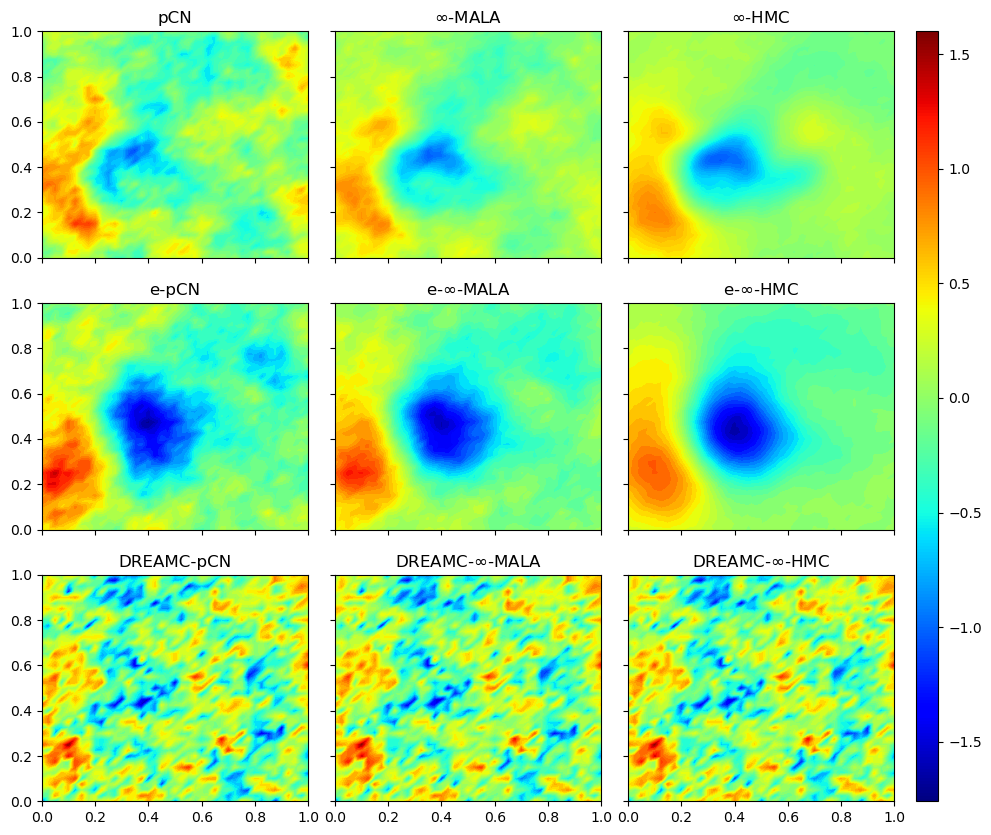}
\caption{Posterior mean estimates of the log-transmissivity field $u(\bx)$.}
\label{fig:elliptic_mcmc_mean}
\end{subfigure}
\begin{subfigure}[b]{1\textwidth}
\includegraphics[width=1\textwidth,height=.46\textwidth]{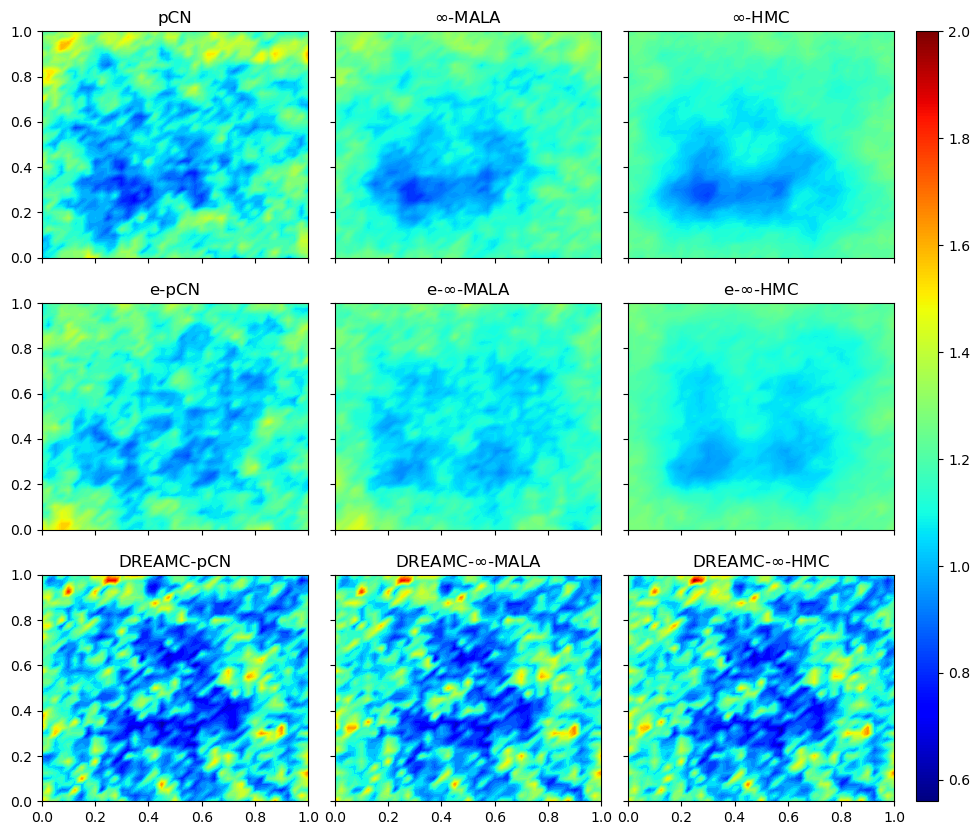}
\caption{Posterior standard deviation estimates of the log-transmissivity field $u(\bx)$.}
\label{fig:elliptic_mcmc_std}
\end{subfigure}
\vspace{-20pt}
\caption{Elliptic inverse problem: Bayesian posterior estimates of the log-transmissivity field $u(\bx)$ based on $5000$ samples by various MCMC algorithms.}
\end{figure}


In the sampling stage, we train AE with the structure illustrated in Figure \ref{fig:ae}. The latent dimension is $d_L=121$ ($11\times 11$) and the sizes of hidden layers between input and latent, between latent and output are linearly interpolated. All the activation functions are chosen as `LeakyReLU($\alpha=2$)'.
Figure \ref{fig:elliptic_reconstruction} plots the original $u^\text{MAP}$ (left), the latent representation $u^\text{MAP}_r= \phi(u^\text{MAP})$ (middle) and the reconstruction $u^{\text{MAP}'}= \psi(u^\text{MAP}_r)$ (right). Even though the latent representation is not very intuitive, the output function (image) decoded from the latent space can be viewed as a `faithful' reconstruction of the original function (image), indicating a sufficiently good AE that compresses and restores information. Therefore, our proposed MCMC algorithms, defined on the latent space, generate samples that can be projected back to the original space without losing too much accuracy in representing the posterior distribution.



\begin{table}[ht]\scriptsize
\centering
\begin{tabular}{l|ccccccc}
  \hline
Method & h $^a$ & AP $^b$ & s/iter $^c$ & ESS(min,med,max) $^d$ & minESS/s $^e$ & spdup $^f$ & PDEsolns $^g$ \\ 
  \hline
pCN & 0.03 & 0.65 & 0.49 & (7.8,28.93,73.19) & 0.0032 & 1.00 & 6001 \\ 
  $\infty$-MALA & 0.15 & 0.61 & 0.56 & (29.21,120.79,214.85) & 0.0105 & 3.30 & 12002 \\ 
  $\infty$-HMC & 0.10 & 0.70 & 1.65 & (547.62,950.63,1411.6) & 0.0663 & 20.82 & 36210 \\ 
    \hline
  e-pCN & 0.05 & 0.60 & 0.02 & (10.07,43.9,93.62) & 0.0879 & 27.60 & 0 \\ 
  e-$\infty$-MALA & 0.15 & 0.67 & 0.03 & (33.23,133.54,227.71) & 0.2037 & 63.95 & 0 \\ 
  e-$\infty$-HMC & 0.10 & 0.77 & 0.07 & (652.54,1118.08,1455.56) & 1.9283 & {\bf 605.47} & 0 \\ 
    \hline
  DREAMC-pCN & 0.10 & 0.67 & 0.02 & (36.78,88.36,141.48) & 0.3027 & 95.03 & 0 \\ 
  DREAMC-$\infty$-MALA & 1.00 & 0.66 & 0.04 & (391.53,782.06,927.08) & 2.0988 & {\bf 659.01} & 0 \\ 
  DREAMC-$\infty$-HMC & 0.60 & 0.64 & 0.11 & (2289.86,3167.03,3702.4) & 4.1720 & {\bf 1309.97} & 0 \\ 
  \hline
\end{tabular}

$^a$ step size\quad
$^b$ acceptance probability\quad 
$^c$ seconds per iteration\quad
$^d$ (minimum, median, maximum) effective sample size\quad

$^e$ minimal ESS per second\quad
$^f$ comparison of minESS/s with pCN as benchmark
$^g$ number of PDE solutions
\caption{Elliptic inverse problem: sampling efficiency of various MCMC algorithms.} 
\label{tab:elliptic}
\end{table}

We compare the performance of algorithms including vanilla pCN, $\infty$-MALA, $\infty$-HMC, their emulative versions, and the corresponding DREAMC algorithms. For each algorithm, we run $6000$ iterations and burn in the first $1000$. For HMC algorithms, we set $I=5$.
We tune the step sizes for each algorithm so that they have similar acceptance rates around $60 \sim 70 \%$.
Figure \ref{fig:elliptic_mcmc_mean} compares their posterior mean estimates, and Figure \ref{fig:elliptic_mcmc_std} compares their estimates of posterior standard deviation.
We can see that emulative MCMC algorithms generate results very close to those by the original MCMC methods. DREAMC algorithms introduce more errors due to the information loss in AE, but still provides estimates that reasonably resemble those generated by the original MCMC algorithms.


Table \ref{tab:elliptic} summarizes the sampling efficiency of various MCMC algorithms measured by minimum effective sample size (ESS) among all parameters normalized by the total time consumption, i.e. minESS/s.
With this standard, emulative $\infty$-HMC and DREAMC $\infty$-MALA achieve more than 600 times speed-up in sampling efficiency and DREAMC $\infty$-HMC attains 3 orders of magnitude improvement compared to the benchmark pCN.
Such comparison focuses on the cost of obtaining uncertainty estimates and does not include the time for training CNN and AE, which is relatively much smaller compared with the overall sampling time.



Figure \ref{fig:elliptic_acf} (Appendix \ref{apx:ext}) shows the traceplots of the potential function (data-misfit) on the left panel and autocorrelation functions on the right panel.
HMC algorithms make distant proposals with least autocorrelation, followed by MALA algorithms and then pCN algorithms with the highest autocorrelation. This is also verified numerically by ESS of parameters (the lower autocorrelation, the higher ESS) in Table \ref{tab:elliptic}.
Note DREAMC $\infty$-MALA has similar autocorrelation as HMC algorithms.
Finally, we plot the Kullback–Leibler (KL) divergence between the posterior and the prior in terms of iteration (upper) and time (lower) respectively in Figure \ref{fig:elliptic_KLt} (Appendix \ref{apx:ext}). 
Among all the MCMC algorithms, emulative MCMC algorithms stabilize such measurement the fastest and attain smaller values for given iterations and time.


\subsection{Advection-Diffusion Inverse Problem}\label{sec:adif}

In the next example, we quantify the uncertainty in the solution of an inverse problem governed by a parabolic PDE within the Bayesian inference framework. 
The underlying PDE is a time-dependent advection-diffusion equation in which we seek to infer an unknown initial condition from spatiotemporal point measurements.

\begin{figure}[t]
\centering
\includegraphics[width=1\textwidth,height=.35\textwidth]{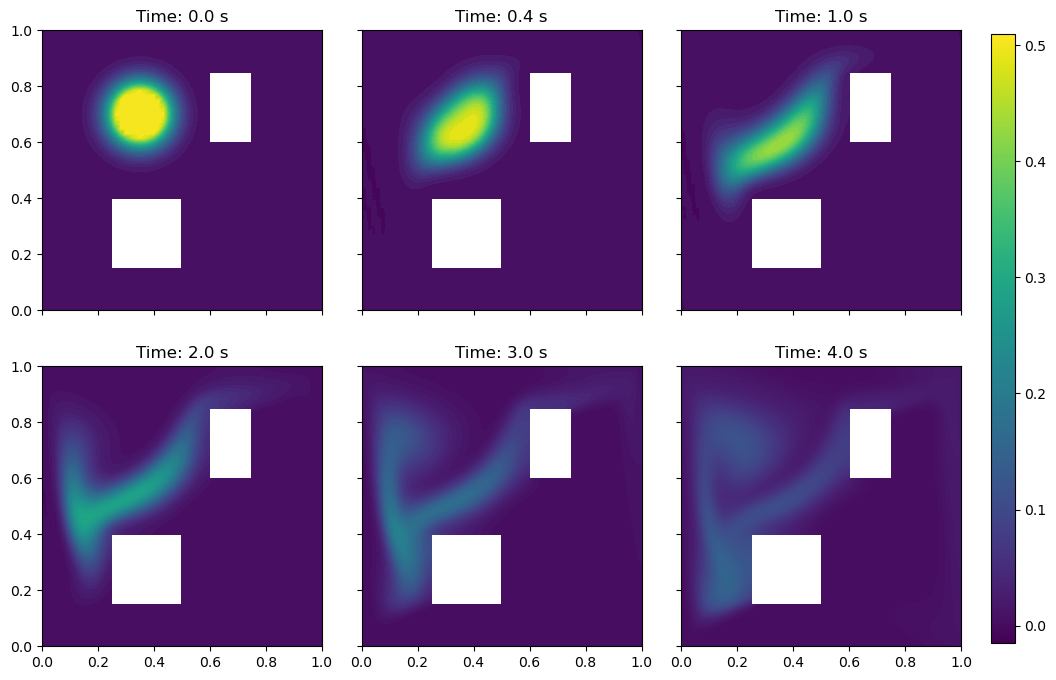}
\caption{Advection-diffusion inverse problem: true initial condition $u_0^\dagger$ (top left), and the solutions $u(\bx, t)$ at different time points.}
\label{fig:time_soln}
\end{figure}

The parameter-to-observable forward mapping $\mG : u_0 \to \mathcal O u$ maps an initial condition $u_0\in L^2(\Omega)$ to pointwise spatiotemporal observations of the concentration field $u(\bx, t)$ through the solution of the following advection-diffusion equation \cite{Petra2011,villa2020}:
\begin{equation}\label{eq:adif}
\begin{aligned}
u_t - \kappa \Delta u + \bv \cdot \nabla u &= 0 \quad in\; \Omega\times (0, T)\\
u(\cdot, 0) &= u_0 \quad in\; \Omega \\
\kappa \nabla u\cdot \vec n &= 0, \quad on\; \pa\Omega\times (0, T)
\end{aligned}
\end{equation}
where $\Omega \subset [0,1]^2$ is a bounded domain shown in Figure \ref{fig:time_soln}, $\kappa >0$ is the diffusion coefficient (set to $10^{-3}$), and $T>0$ is the final time.
The velocity field $\bv$ is computed by solving the following steady-state Navier-Stokes equation with the side walls driving the flow \cite{Petra2011}:
\begin{equation}\label{eq:adif_v}
\begin{aligned}
-\frac{1}{\mathrm{Re}} \Delta \bv + \nabla q + \bv \cdot \nabla \bv &= 0 \quad in\; \Omega\\
\nabla \cdot \bv &= 0 \quad in\; \Omega \\
\bv &={\bf g}, \quad on\; \pa\Omega
\end{aligned}
\end{equation}
Here, $q$ is the pressure, $\mathrm{Re}$ is the Reynolds number, which is set to 100 in this example. The Dirichlet boundary data ${\bf g}\in \mbR^2$ is given by ${\bf g}={\bf e}_2=(0,1)$ on the left wall of the domain, ${\bf g}=-{\bf e}_2$ on the right wall, and ${\bf g}={\bf 0}$ everywhere else.

\begin{figure}[t]
\begin{subfigure}[b]{1\textwidth}
\includegraphics[width=1\textwidth,height=.25\textwidth]{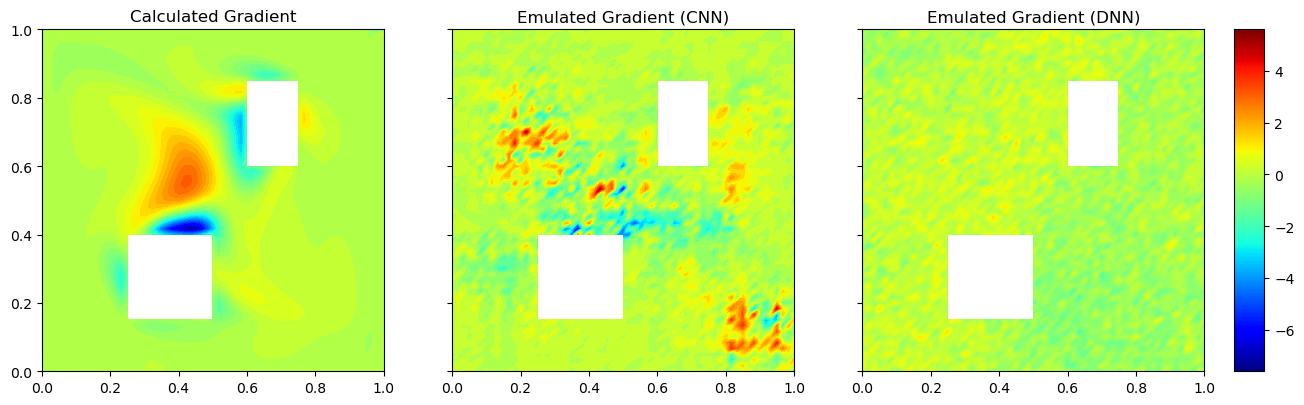}
\caption{CNN (middle) and DNN (right) emulation ($\mG^e: \mbR^{3413}\to\mbR^{1280}$) extracting gradients $D\Phi^e(u^\text{MAP})$ compared with the true gradient $D\Phi(u^\text{MAP})$ (left).}
\label{fig:adif_extrctgrad}
\end{subfigure}
\begin{subfigure}[b]{1\textwidth}
\includegraphics[width=1\textwidth,height=.25\textwidth]{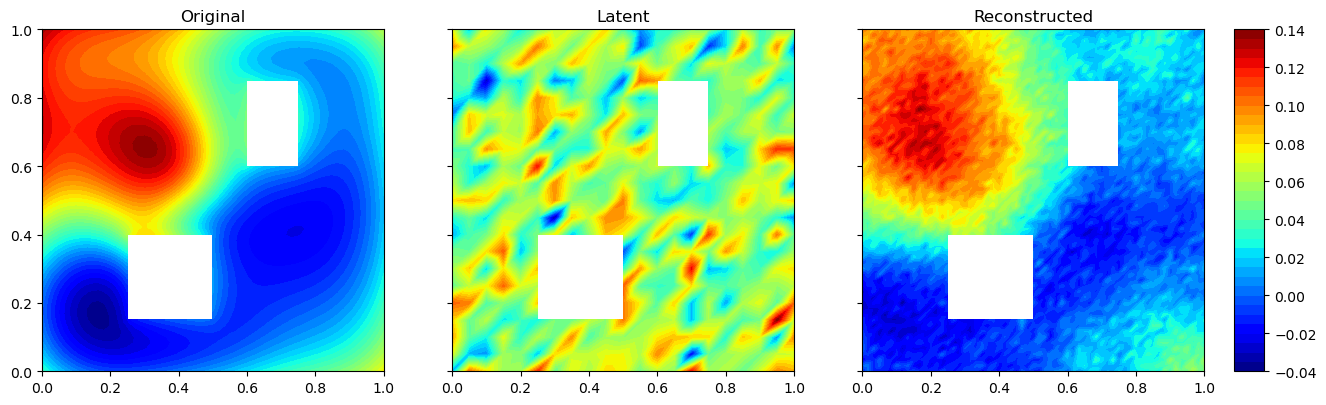}
\caption{AE compressing the original function $u^\text{MAP}$ (left) into latent space $u^\text{MAP}_r$ (middle) and reconstructing it in the original space $u^{\text{MAP}'}$ (right).}
\label{fig:adif_reconstruction}
\end{subfigure}
\vspace{-20pt}
\caption{Advection-diffusion inverse problem: outputs by NNs viewed as 2d images.}
\end{figure}

We set the true initial condition $u_0^\dagger = 0.5\wedge \exp\{-100[(x_1-0.35)^2+(x_2-0.7)^2]\}$, illustrated in the top left panel of Figure \ref{fig:time_soln}, which also shows a few snapshots of solutions $u$ at other time points on a regular grid mesh of size $61\times 61$.
To obtain spatiotemporal observations, we collect solutions $u(\bx, t)$ solved on a refined mesh at 80 selected locations $\{\bx_i\}_{i=1}^{80}$ (Figure \ref{fig:spatiotemporal_obs}) across 16 time points $\{t_j\}_{j=1}^{16}$ evenly distributed between 1 and 4 seconds (thus denoted as $\mathcal O u$) and inject some Gaussian noise $\mN(0, \sigma^2_\eta)$ such that the relative noise standard deviation is 0.5 ($\sigma_\eta/\max \mathcal O u = 0.5$):
\begin{equation*}
y = \mathcal O u(\bx, t) + \eta, \quad \eta \sim \mathcal N(0, \sigma_\eta^2 I_{1280})
\end{equation*}
In the Bayesian setting, we adopt the following GP prior with the covariance kernel $\mC$ defined through the Laplace operator $\Delta$:
\begin{equation*}
u \sim \mu_0 = \mN(0, \mC), \quad \mC = (\delta \mI -\gamma \Delta )^{-2}
\end{equation*}
where $\delta$ governs the variance of the prior and $\gamma/\delta$ controls the correlation length. We set $\gamma=2$ and $\delta=10$ in this example.

\begin{figure}[tbp]
\begin{subfigure}[b]{1\textwidth}
\includegraphics[width=1\textwidth,height=.46\textwidth]{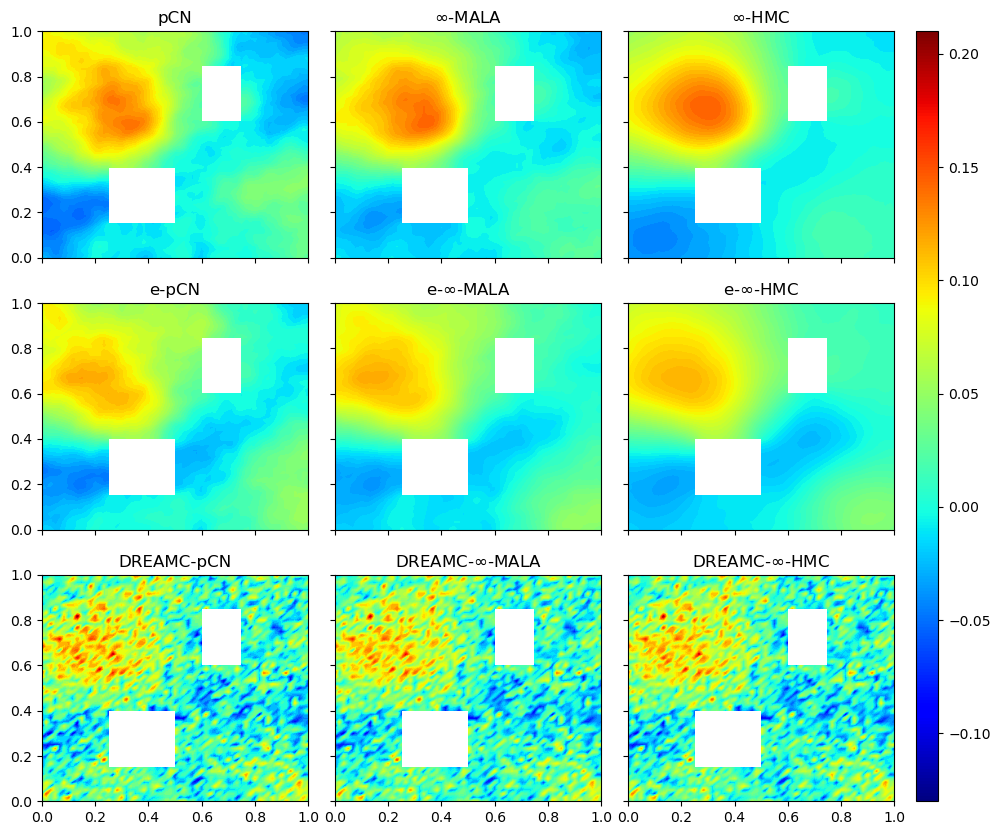}
\caption{Posterior mean estimates of the initial concentration field $u(\bx)$.}
\label{fig:adif_mcmc_mean}
\end{subfigure}
\begin{subfigure}[b]{1\textwidth}
\includegraphics[width=1\textwidth,height=.46\textwidth]{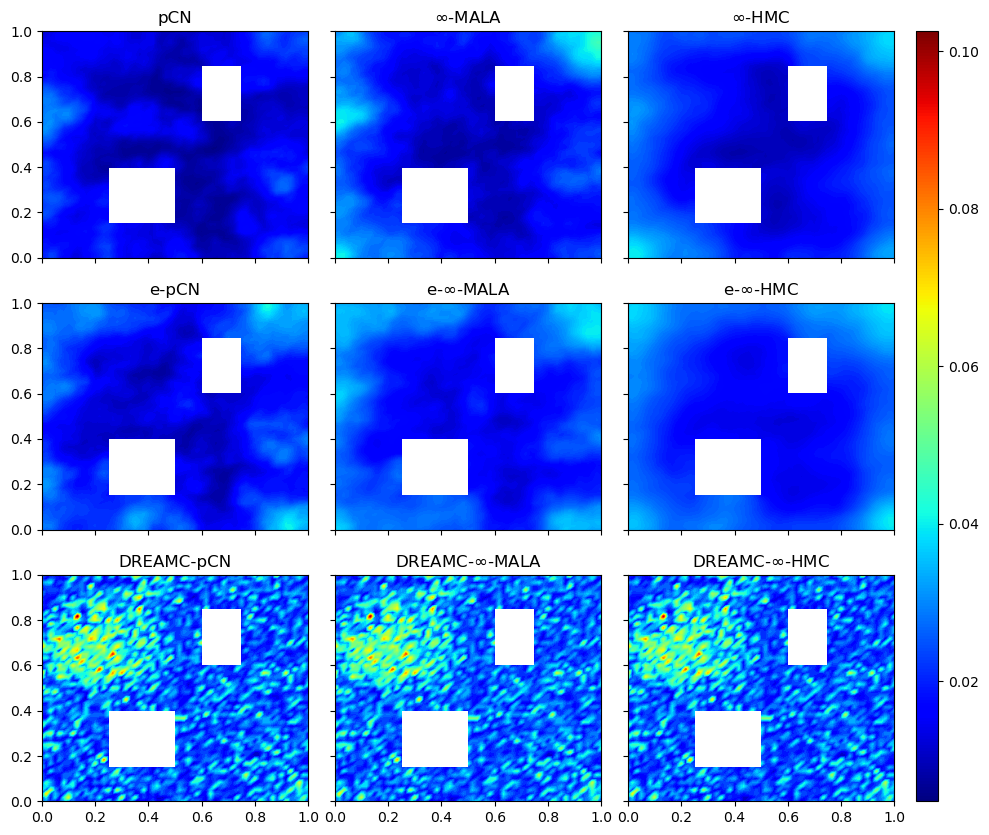}
\caption{Posterior standard deviation estimates of the initial concentration field $u(\bx)$.}
\label{fig:adif_mcmc_std}
\end{subfigure}
\vspace{-20pt}
\caption{Advection-diffusion inverse problem: Bayesian posterior estimates of the initial concentration field $u(\bx)$ based on $5000$ samples by various MCMC algorithms.}
\end{figure}

The Bayesian inverse problem involves obtaining an estimate of the initial condition $u_0$ and quantifying its uncertainty based on the $80\times 16$ spatiotemporal observations $y\in\mbR^{1280}$.
For the notational convenience, we still denote $u_0(\bx)$ as $u(\bx)$ when it is not confused with the general concentration field $u(\bx, t)$.
The Bayesian UQ in this example is especially challenging not only because of its large dimensionality (3413) of spatially discretized $u$ (Lagrange degree 1) at each time $t$, but also due to the spatiotemporal interactions in these observations \cite{cressie2011}.
We follow the CES framework as in Section \ref{sec:dream}. 
In the calibration stage, we collect $\{u_n^{(j)}, \mG(u_n^{(j)})\}_{j=1,n=1}^{J,N}$ by running EKS with the ensemble size $J=500$ for $N=10$ iterations.
For the emulation, we train DNN and CNN with the same $3:1$ splitting of these 5000 ensembles for training and testing.
The DNN has 5 layers with activation function `LeakyReLU($\alpha=0.01$)' for the hidden layers and `linear' activation for the output layer and $25\%$ nodes dropped out.
The structure of CNN is illustrated in Figure \ref{fig:cnn} with 4 filters in the last convolution layer, activation `LeakyReLU($\alpha=0.2$)'  for the convolution layers, `PReLU' activation for the latent layer (dimension 1024), and `linear' activation for the output layer. The trained CNN has drop out rate of $50\%$ on all its nodes.
Figure \ref{fig:adif_extrctgrad} compares the true gradient function $D\Phi(u^\text{MAP})$ (left panel) and its emulations $D\Phi^e(u_\text{MAP})$ (middle and right panels) as in Equation \eqref{eq:dpotential_emu} extracted from two types of NNs. As before, we can see better extracted gradient output by CNN as an approximation to the true gradient compared with DNN.
Due to the large dimensionality of inputs and outputs ($\mG^e: \mbR^{3413}\to\mbR^{1280}$) and memory requirement, GP (implemented in \texttt{GPflow}) failed to fit and output gradient extraction.


In the sampling stage, we adopt AE with the same structure as in Figure \ref{fig:ae}, the latent dimension $d_L=417$ (degrees of freedom on submesh $21\times 21$), and the activation functions chosen as `elu'.
Figure \ref{fig:adif_reconstruction} plots the original $u^\text{MAP}$ (left), the latent representation $u^\text{MAP}_r= \phi(u^\text{MAP})$ (middle) and the reconstruction $u^{\text{MAP}'}= \psi(u^\text{MAP}_r)$ (right). 
Again we can see a `faithful' reconstruction of the original function (image) by AE even though the latent representation is not very intuitive.



\begin{table}[ht]\scriptsize
\centering
\begin{tabular}{l|ccccccc}
  \hline
Method & h $^a$ & AP $^b$ & s/iter $^c$ & ESS(min,med,max) $^d$ & minESS/s $^e$ & spdup $^f$ & PDEsolns $^g$ \\ 
  \hline
pCN & 0.001 & 0.69 & 0.03 & (3.16,6.37,40.7) & 0.0222 & 1.00 & 6001 \\ 
  $\infty$-MALA & 0.005 & 0.68 & 0.06 & (3.78,11.6,51.5) & 0.0122 & 0.55 & 12002 \\ 
  $\infty$-HMC & 0.005 & 0.78 & 0.12 & (31.55,83.54,240.34) & 0.0507 & 2.29 & 35872 \\ 
  \hline
  e-pCN & 0.002 & 0.69 & 0.02 & (3.33,7.19,58.2) & 0.0324 & 1.46 & 0 \\ 
  e-$\infty$-MALA & 0.008 & 0.72 & 0.05 & (4.28,14.3,62) & 0.0157 & 0.71 & 0 \\ 
  e-$\infty$-HMC & 0.008 & 0.72 & 0.11 & (25.41,113.11,270.79) & 0.0475 & 2.14 & 0 \\ 
  \hline
  DREAMC-pCN & 0.020 & 0.68 & 0.02 & (8.88,16.99,53.35) & 0.0727 & {\bf 3.28} & 0 \\ 
  DREAMC-$\infty$-MALA & 0.100 & 0.83 & 0.06 & (37.65,66.58,157.09) & 0.1310 & {\bf 5.91} & 0 \\ 
  DREAMC-$\infty$-HMC & 0.100 & 0.72 & 0.17 & (564.12,866.72,1292.11) & 0.6791 & {\bf 30.64} & 0 \\ 
   \hline
\end{tabular}

$^a$ step size\quad
$^b$ acceptance probability\quad 
$^c$ seconds per iteration\quad
$^d$ (minimum, median, maximum) effective sample size\quad

$^e$ minimal ESS per second\quad
$^f$ comparison of minESS/s with pCN as benchmark
$^g$ number of PDE solutions
\caption{Advection-diffusion inverse problem: sampling efficiency of MCMC algorithms.} 
\label{tab:adif}
\end{table}

We compare the performance of $\infty$-MCMC algorithms (pCN, $\infty$-MALA, $\infty$-HMC), their emulative versions, and the corresponding DREAMC algorithms. For each algorithm, we run $6000$ iterations and burn in the first $1000$. For HMC algorithms, we set $I=5$.
We tune the step sizes for each algorithm so that they have similar acceptance rates around $60 \sim 70 \%$.
Figure \ref{fig:adif_mcmc_mean} compares their posterior mean estimates and Figure \ref{fig:adif_mcmc_std} compares their estimates of posterior standard deviation.
We can see that emulative MCMC algorithms generate similar results as the original MCMC methods. 
DREAMC algorithms yield estimates close enough to those by the original MCMC. Although there are some deviations in the uncertainty estimates, the results by DREAMC algorithms are significantly better than those by ensemble Kalman methods, which severely underestimate the posterior standard deviations (See Figure \ref{fig:adif_std500}).

Table \ref{tab:adif} compares the sampling efficiency of various MCMC algorithms measured by minESS/s.
Three most efficient sampling algorithms are all DREAMC algorithms. DREAMC $\infty$-HMC attains up to 30 times speed up compared to the benchmark pCN.
Considering the complexity of this inverse problem with spatiotemporal observations, this is a significant achievement.
Again, we exclude the training time of CNN and AE from the comparison since it is rather negligible compared with the overall sampling time.

In Appendix \ref{apx:ext},
Figure \ref{fig:adif_acf} verifies DREAMC $\infty$-HMC is the most efficient MCMC algorithm that has the smallest autocorrelation shown on the right panel. It is followed by other HMC algorithms and DREAMC $\infty$-MAMA which is even better than $\infty$-HMC.
Figure \ref{fig:adif_KLt} plots the KL divergence between the posterior and the prior in terms of iteration (upper) and time (lower) respectively. As we can see, $\infty$-HMC converges the fastest.

%


\section{Conclusion}\label{sec:con}

In this paper, we have proposed a new framework to scale up Bayesian UQ for inverse problems. More specifically, we use CNN -- a regularized neural network, which is a powerful tool for image recognition and amenable to inverse problems if we treat the discretized parameter function as an input image. This way, CNN is capable of learning spatial features. In addition, the resulting algorithm has low computational complexity and is robust: as seen in Figure \ref{fig:gp_cnn}, the performance of CNN as an emulator is relatively stable across different training sizes. If larger training size is required for certain problems, we could train CNN adaptively as more samples are collected from the parameter space \cite{shahbaba2019}. We have adopted AE to further reduce the dimension of the parameter space and speed up the sampling process. Overall, by utilizing different techniques based on neural networks, we have been able to scale up Bayesian UQ up to thousands of dimensions.

In the current framework, we rely on regular grid mesh to facilitate the CNN training -- a discretized function over a 2d mesh needs to be converted to a matrix of image pixels.
Such a limitation can be alleviated by using mesh CNN \cite{Hanocka_2019}, which could train CNN directly on irregular mesh; for example, triangular mesh has been extensively used for solving PDE.
This will extend our methodology and further enhances its utility.


The standard AE used in our proposed framework and the corresponding latent projection by dense layers might not be the optimal choices. Alternatively, we could use convolutional AE (CAE) \cite{Guo_2017}, which generates more recognizable latent representation as illustrated in Figure \ref{fig:elliptic_reconstruction_cae}.
In this case, the latent parameter can be interpreted as a representation of the original function on a coarser mesh.
What is more, we could modify the loss function to adaptively learn the dimensionality of intrinsic latent space.

There are spatiotemporal data in some inverse problems (e.g., advection-diffusion equation). In such cases, we could model the temporal pattern of observations in the emulation using some recurrent neural networks (RNN) \cite{Rumelhart_1986}, e.g., long short-term memory (LSTM) \cite{Hochreiter_1997}. We can then build a `CNN-RNN' emulator with convolutional layer for function (image) inputs and RNN layer for multivariate time series outputs. While we have obtained promising preliminary results (See Figure \ref{fig:cnn_cnnrnn}), we intend to pursue this idea in a follow-up paper.
Lastly, future research could involve replacing the MCMC algorithms in the sampling step with some recently proposed information gradient flow methods \cite{Wang_2021,wang2020} and comparing the performance with the current DREAMC approach.


\newpage
\bibliographystyle{siamplain}
\bibliography{references}

\end{document}


\maketitle

\section{Proof of Theorem \ref{thm:cnn_err}}\label{apx:cnn_err}

\begin{thm*}[3.1]
Let $2\leq s\leq d$ and $\Omega\subset [-1,1]^d$. Assume $\mG_j\in H^r(\mbR^d)$ with $r\geq 1$ such that $v_{\mG_j,2}:=\int_{\mbR^d}\Vert \omega\Vert_1^2|\widehat \mG_j(\omega)| d\omega<\infty$ for $j=1,\cdots, m$.
If $K\geq 2d/(s-1)$, then there exist $\mG^e$ by CNN with ReLU activation function such that
\begin{equation}\label{eq: cnn_errorbound2}
\Vert \Phi - \Phi^e\Vert_{H^1(\Omega)} \leq c v_{\mG,2} \sqrt{\log K} K^{-\half - \frac{1}{2d}}
\end{equation}
where we have $\Vert\Phi\Vert_{H^1(\Omega)} = \left(\Vert\Phi\Vert_{L^2(\Omega)}^2+\Vert D\Phi\Vert_{L^2(\Omega)}^2\right)^\half$, $c$ is an absolute constant, and $v_{\mG,2}=\max_{1\leq j\leq m} v_{\mG_j,2}$.
\end{thm*}

\begin{proof}
Note because we have
\begin{align*}
|\Phi(u) - \Phi^e(u)| \leq |\langle \mG(u) - \mG^e(u), y - \mG(u) \rangle_\Gamma | + |\langle y - \mG^e(u), \mG(u) - \mG^e(u)\rangle_\Gamma | \\
|D\Phi(u) - D\Phi^e(u)| \leq |\langle \mG(u) - \mG^e(u), D\mG(u) \rangle_\Gamma | + |\langle y - \mG^e(u), D\mG(u) - D\mG^e(u)\rangle_\Gamma |
\end{align*}
it suffices to prove
\begin{equation}\label{eq:suff}
\Vert \mG_j - \mG_j^e\Vert_{H^1(\Omega)} \leq c_j \Vert \mG_j\Vert \sqrt{\log K} K^{-\half - \frac{1}{2d}}, \quad j=1,\cdots, m
\end{equation}
Let $K^*$ be the integer part of $\frac{(s-1)K}{d}-1$, i.e. $K^*=\left[\frac{(s-1)K}{d}\right]-1\geq 1$.
For each $j=1,\cdots, m$, there exists a linear combination of ramp ridge functions of the following form by \cite[Theorem 2]{Klusowski_2018}:
\begin{equation}\label{eq: cnn_construction}
\mG_j^e(u) = \mG_j(0) + D \mG_j(0) \cdot u + \frac{v}{K^*} \sum_{k=1}^{K^*} \beta_k (\alpha_k \cdot u - t_k)_+
\end{equation}
with $\beta_k\in[-1,1]$, $\Vert\alpha_k\Vert_1=1$, $t_k\in[0,1]$ and $|v|\leq 2 v_{\mG_j,2}:=\int_{\mbR^d}\Vert \omega\Vert_1^2|\widehat \mG_j(\omega)| d\omega \leq c_{d,r} \Vert \mG_j\Vert$ such that
\begin{equation}
\Vert \mG_j - \mG_j^e\Vert_{L^\infty([-1,1]^d)} \leq c_0 v_{\mG_j,2} (\sqrt{\log K^*} \vee \sqrt d \,) (K^*)^{-\half - \frac{1}{d}}
\end{equation}
\cite[Theorem 2]{ZHOU_2020} constructs weights $W$ and biases $b$ of a CNN that has output of the form in Equation \eqref{eq: cnn_construction}. Therefore,
\begin{equation}\label{eq:err1}
\Vert \mG_j - \mG_j^e\Vert_{L^2(\Omega)} \leq C \Vert \mG_j - \mG_j^e\Vert_{L^\infty(\Omega)} \leq c_j \Vert \mG_j\Vert \sqrt{\log K} K^{-\half - \frac{1}{d}}, \quad j=1,\cdots, m
\end{equation}

Now we take derivative on both sides of \eqref{eq: cnn_construction} to get
\begin{equation}\label{eq: cnn_gradient}
D\mG_j^e(u) = D \mG_j(0) + \frac{v}{K} \sum_{k=1}^K \alpha_k \beta_k H(\alpha_k \cdot u - t_k)
\end{equation}
where $H(x) = I(x\geq 0)$ is the Heaviside function.
For any $i=1,\cdots, d$, we have $v_{D_i\mG_j,1}:=\int_{\mbR^d}\Vert \omega\Vert_1 |\widehat{D_i\mG}_j(\omega)| d\omega \leq C \int_{\mbR^d}\Vert \omega\Vert_1 |\omega_i| |\widehat{\mG}_j(\omega)| d\omega\leq C v_{\mG_j,2}$.
Therefore, by Theorem 3 of \cite{MAKOVOZ_1996} we have
\begin{equation}\label{eq:err2}
\Vert D_i\mG_j - D_i\mG_j^e\Vert_{L^2([-1,1]^d)} \leq c_0' v_{\mG_j,2} K^{-\half - \frac{1}{2d}}
\end{equation}
Inequality \eqref{eq:err1} and inequality \eqref{eq:err2} yield error bound \eqref{eq:suff} thus complete the proof.
\end{proof}

\section{Algorithms}\label{apx:alg}

We give details of DREAMC-$\infty$-HMC in Algorithm \ref{alg:DRe-infHMC}. 
When $I=1$,  DREAMC-$\infty$-HMC reduces to DREAMC-$\infty$-MALA. Further, if $\alpha=0$, the algorithm becomes DREAMC-pCN.

\begin{algorithm}[ht]
\caption{Dimension Reduced Emulative Autoencoder $\infty$-dimensional HMC (DREAMC-$\infty$-HMC)}
\label{alg:DRe-infHMC}
\centering
\begin{algorithmic}[1]
\REQUIRE Collect $NJ$ samples $\{u_n^{(j)}, \mG(u_n^{(j)})\}_{j,n}$ from EKI or EKS procedure; whiten coordinates $\{\tilde u_n^{(j)} = \mC^{-\half} u_n^{(j)}\}_{j,n}$.
\REQUIRE Build an emulator of the forward mapping $\mG^e$ based on $\{\tilde u_n^{(j)}, \mG^{(j)}\}_{j,n}$ (and extract $D\mG^e$) using CNN; train an AE $(\phi,\psi)$ based on $\{\tilde u_n^{(j)}\}_{j,n}$;
\STATE Initialize current state $\tilde u_0$ and project it to the latent space by $\tilde u_{L,0}=\phi(\tilde u_0)$
\STATE Sample velocity $\tilde v_{L,0}\sim \mN(0,I_{d_L})$
\STATE Calculate current energy $E_0=\Phi^e_r(\tilde u_{L,0}) - \frac{\alpha^2\eps^2}{8} \Vert D\Phi^e_r(\tilde u_{L,0})\Vert^2 + \log\det (d\phi(\tilde u_0))$
\FOR{ $i = 0$ to $I-1$}
\STATE Run $\Psi_\eps: (\tilde u_{L,i},\tilde v_{L,i})\mapsto (\tilde u_{L,i+1}, \tilde v_{L,i+1})$ according to \eqref{eq:mHDdiscret_white}.
\STATE Update the energy $E_0 \gets E_0 + \frac{\alpha\eps}{2} (\langle \tilde v_{L,i}, D\Phi^e_r(\tilde u_{L,i}) \rangle + \langle \tilde v_{L,i+1}, D\Phi^e_r(\tilde u_{L,i+1}) \rangle)$
\ENDFOR
\STATE Calculate new energy $E_1=\Phi^e_r(\tilde u_{L,I}) - \frac{\alpha^2\eps^2}{8} \Vert D\Phi^e_r(\tilde u_{L,I})\Vert^2 - \log\det (d\psi(\tilde u_{L,I}))$
\STATE Calculate acceptance probability $a=\exp(-E_1+E_0)$
\STATE Accept $\tilde u_{L,I}$ with probability $a$ for the next state $\tilde u'_L$ or set $\tilde u'_L=\tilde u_{L,0}$ in the latent space.
\STATE Record the next state $u'=\mC^\half \psi(\tilde u'_L)$ in the original space.
\end{algorithmic}
\end{algorithm}

\section{Extended Results}\label{apx:ext}

Below we include some figures for extended results.
\begin{figure}[t]
\begin{subfigure}[b]{1\textwidth}
\includegraphics[width=1\textwidth,height=.25\textwidth]{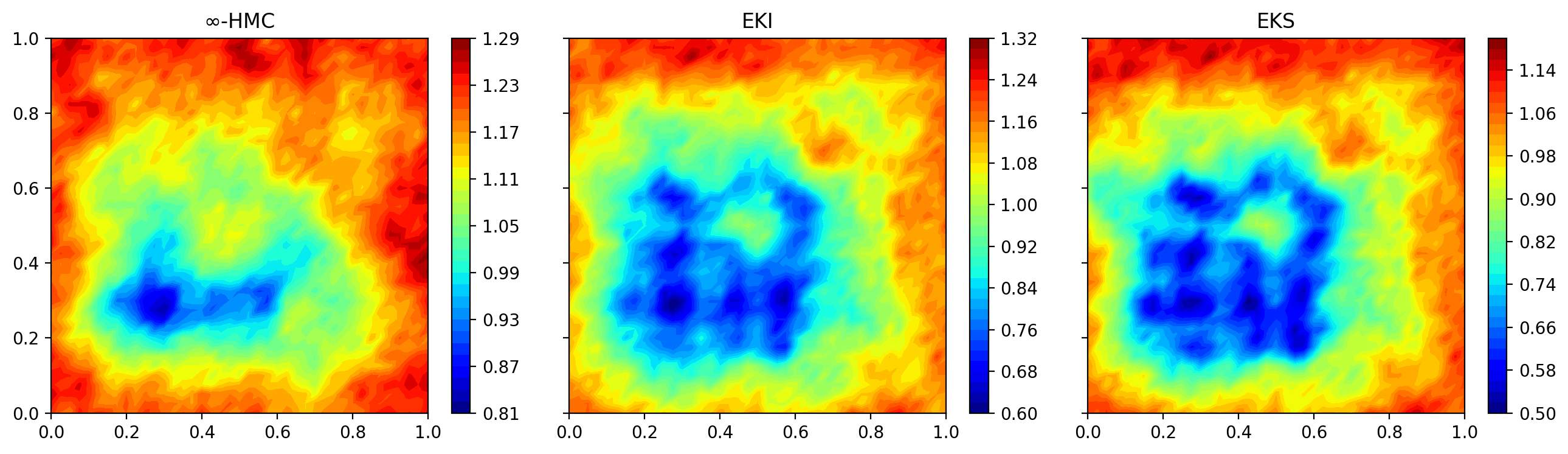}
\caption{Elliptic inverse problem: posterior standard deviation estimated by MCMC (left) and ensemble Kalman methods (right two) with 500 ensembles.}
\label{fig:elliptic_std500}
\end{subfigure}
\begin{subfigure}[b]{1\textwidth}
\includegraphics[width=1\textwidth,height=.25\textwidth]{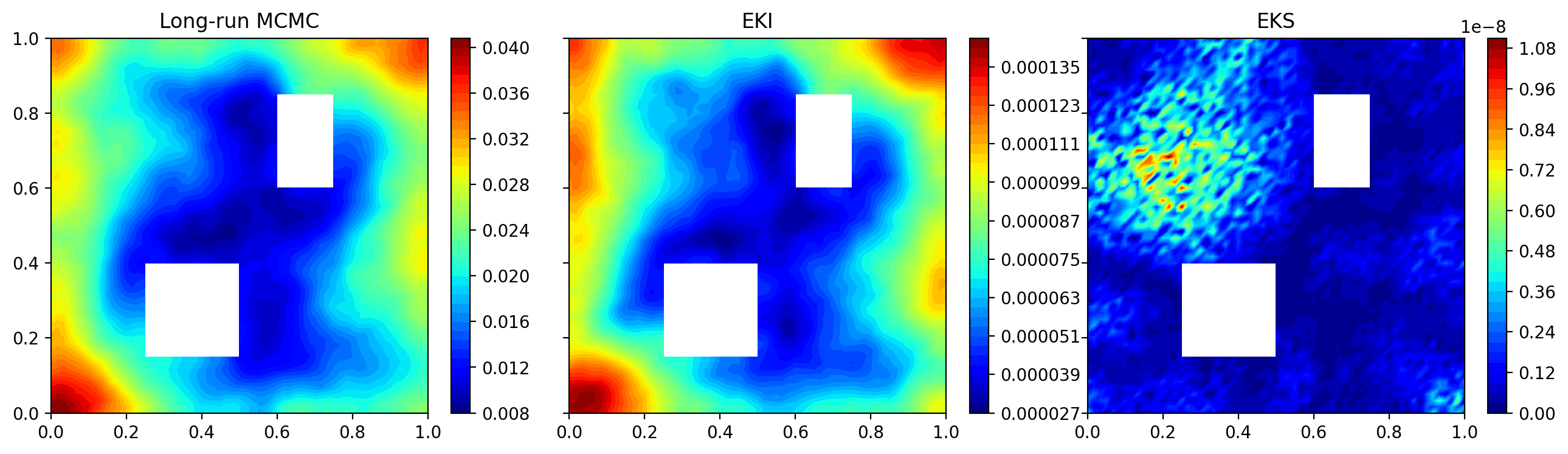}
\caption{Advection-diffusion inverse problem: posterior standard deviation estimated by MCMC (left) and ensemble Kalman methods (right two) with 500 ensembles.}
\label{fig:adif_std500}
\end{subfigure}
\vspace{-20pt}
\caption{Comparing the estimation of posterior standard deviation.}
\end{figure}
In general, ensembles tend to collapse and ensemble Kalman methods cannot provide accurate UQ with finite ensemble size.
Compared to Figure \ref{fig:ensbl_std} with $J=100$ ensembles in the elliptic inverse problem, 
Figures \ref{fig:elliptic_std500} shows that more ensembles (500) might help produce better posterior standard deviation estimates (right two) comparable to gold-standard MCMC (left).
This of course depends on the problem at hand. As Figure \ref{fig:adif_std500} shows both ensemble methods (right two) with 500 ensembles still significantly underestimate the posterior standard deviation in the advection-diffusion inverse problem.

\begin{figure}[ht]
	\centering
	\includegraphics[height=.25\textwidth,width=1\textwidth]{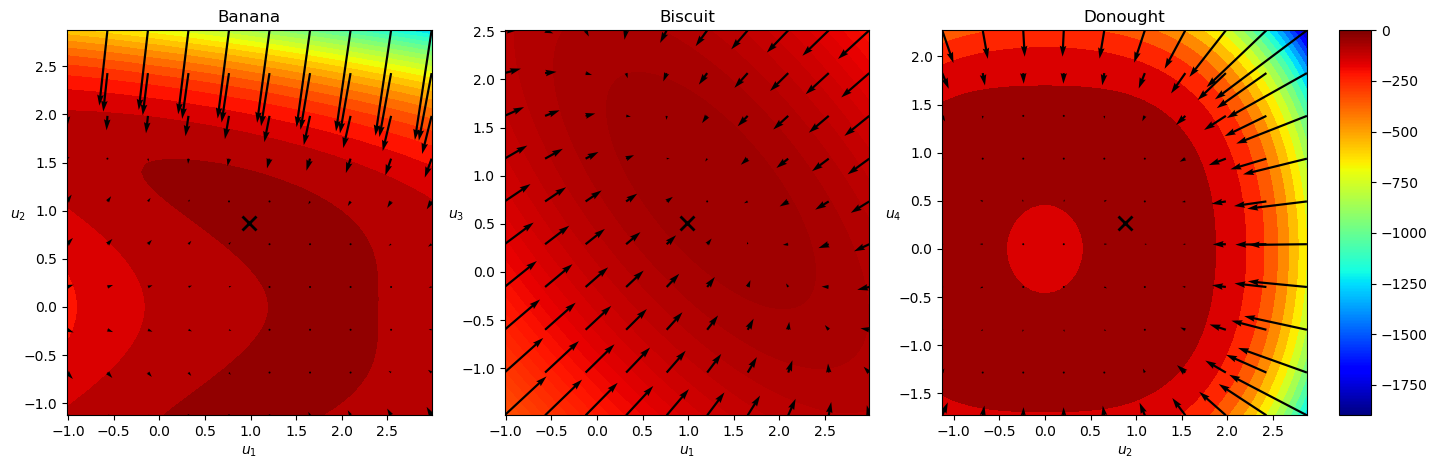}
	\vspace{-10pt}
	\caption{The Banana-Biscuit-Doughnut (BBD) distribution.}
	\label{fig:bbd}
\end{figure}

The non-linear BBD inverse problem involves complex posterior distribution that serves as a good benchmark for UQ methods.
The name comes from the geometric structure of the pairwise posterior distributions resembling a banana in (1,2) dimension (left), a biscuit in (1,3) dimension (middle), and a doughnut in (2,4) dimension (right) as shown in Figure \ref{fig:bbd}.

\begin{figure}[ht]
\begin{subfigure}[b]{1\textwidth}
\includegraphics[width=1\textwidth,height=.3\textwidth]{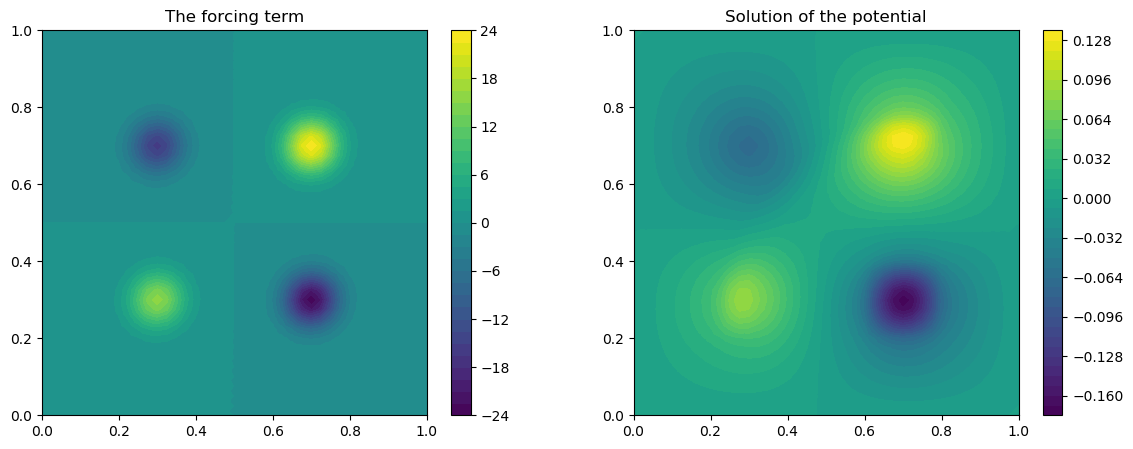}
\caption{Forcing field $f(\bx)$ (left), and the solution $p(\bx)$ with true transmissivity field $k^\dagger(\bx)$ (right).}
\label{fig:force_soln}
\end{subfigure}
\begin{subfigure}[b]{1\textwidth}
\includegraphics[width=1\textwidth,height=.35\textwidth]{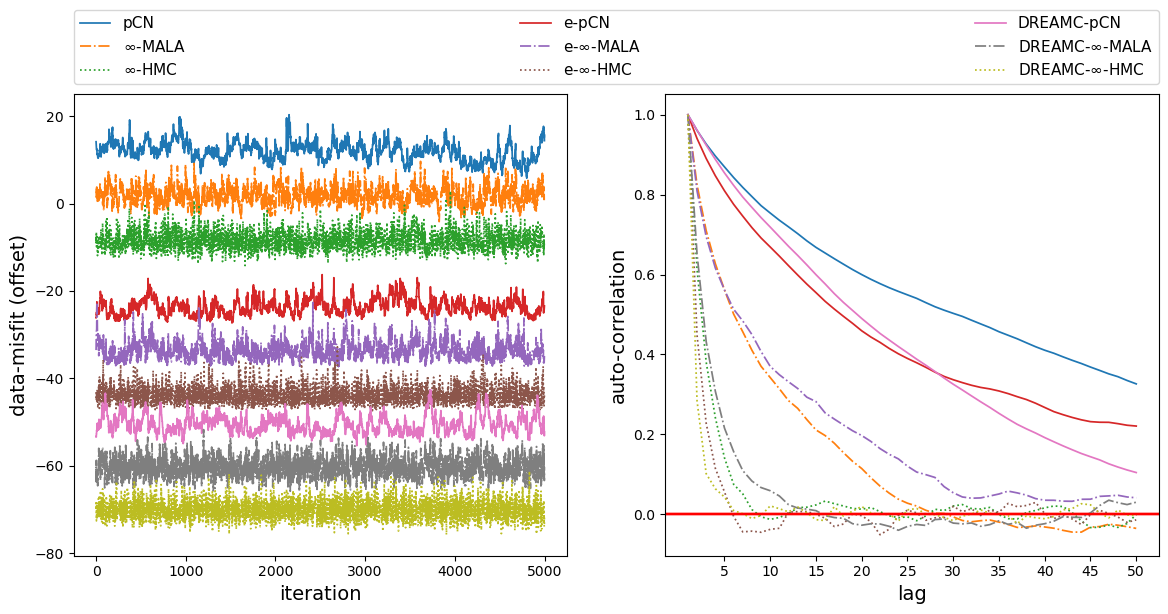}
\caption{The trace plots of data-misfit function evaluated with each sample (left, values have been offset to be better compared with) and the auto-correlation of data-misfits as a function of lag (right).}
\label{fig:elliptic_acf}
\end{subfigure}
\begin{subfigure}[b]{1\textwidth}
\includegraphics[width=1\textwidth,height=.35\textwidth]{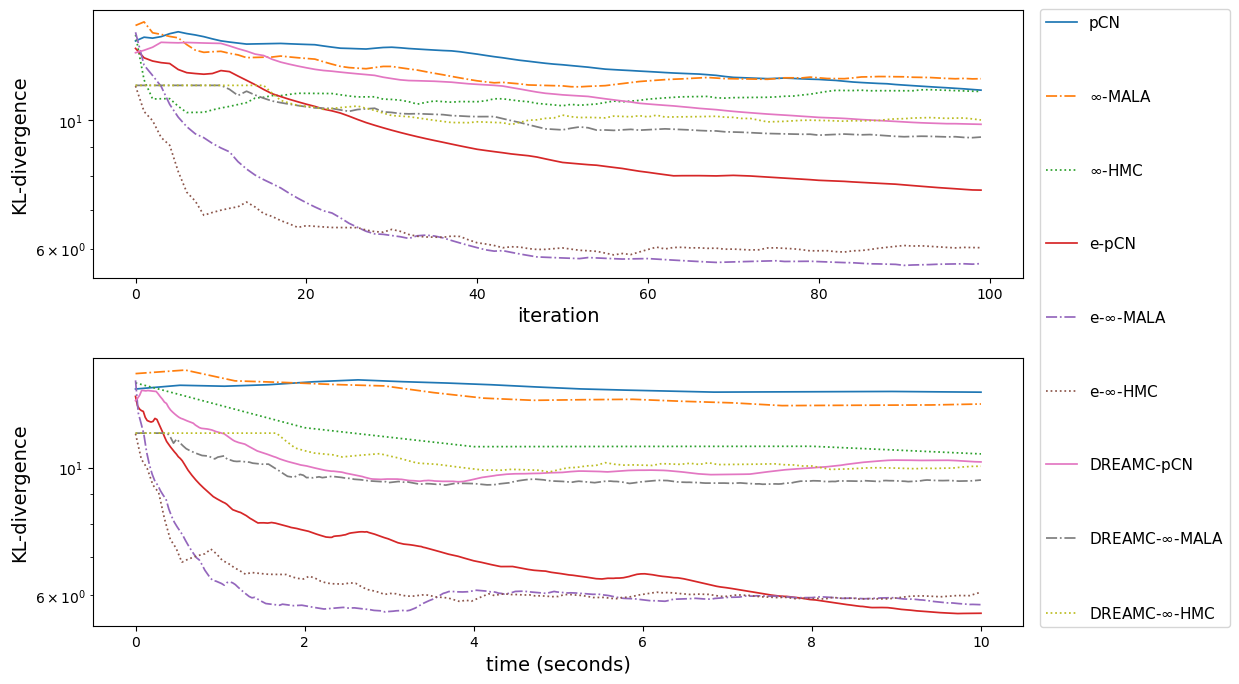}
\caption{The KL divergence between the posterior and the prior as a function of iteration (upper) and time (lower) respectively.}
\label{fig:elliptic_KLt}
\end{subfigure}
\vspace{-20pt}
\caption{Elliptic inverse problem.}
\end{figure}

For the elliptic inverse problem (Section \ref{sec:elliptic}), Figure \ref{fig:elliptic_acf} (Appendix \ref{apx:ext}) shows the traceplots of the potential function (data-misfit) on the left panel and autocorrelation functions on the right panel.
HMC algorithms make distant proposals with the least level of autocorrelation, followed by MALA algorithms, and then by pCN algorithms with the highest level of autocorrelation. This is also verified numerically by ESS of parameters (the lower autocorrelation, the higher ESS) in Table \ref{tab:elliptic}.
Note DREAMC $\infty$-MALA has similar autocorrelation as HMC algorithms.
Finally, we plot the Kullback–Leibler (KL) divergence between the posterior and the prior in terms of iteration (top panel) and time (bottom panel) respectively in Figure \ref{fig:elliptic_KLt} (Appendix \ref{apx:ext}). 
Among all the MCMC algorithms, emulative MCMC algorithms stabilize these measurements the fastest and attain smaller values for given iterations and time.

\begin{figure}[ht]
\begin{subfigure}[b]{1\textwidth}
\includegraphics[width=1\textwidth,height=.35\textwidth]{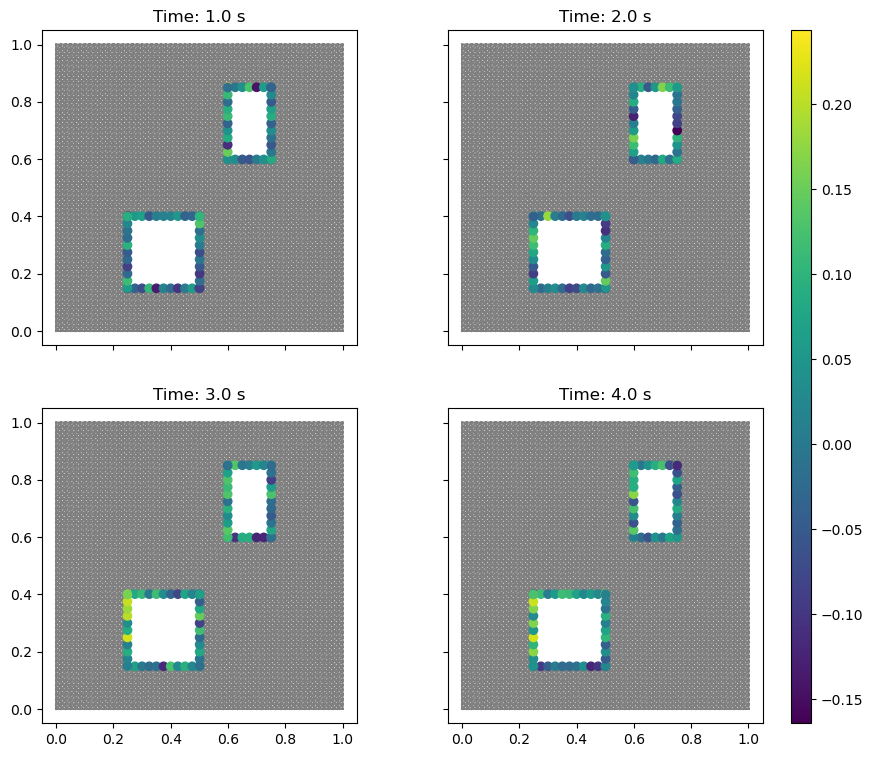}
\caption{Advection-diffusion inverse problem: spatiotemporal observations at $80$ selected locations indicated by circles across different time points, with color indicating their values.}
\label{fig:spatiotemporal_obs}
\end{subfigure}
\begin{subfigure}[b]{1\textwidth}
\includegraphics[width=1\textwidth,height=.35\textwidth]{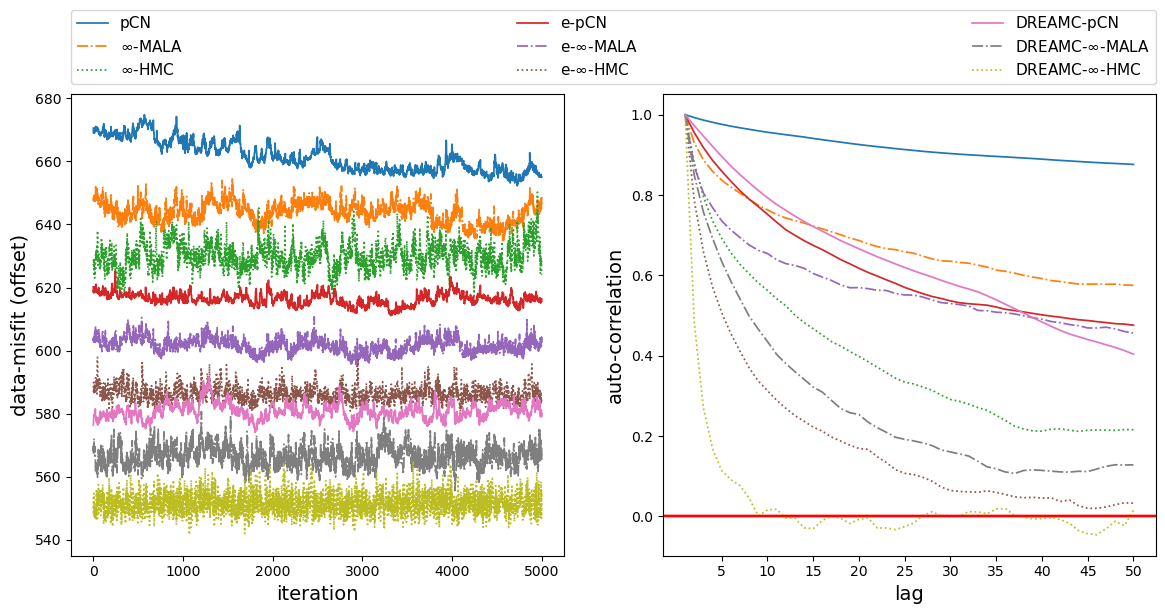}
\caption{The trace plots of data-misfit function evaluated with each sample (left, values have been offset to be better compared with) and the auto-correlation of data-misfits as a function of lag (right).}
\label{fig:adif_acf}
\end{subfigure}
\begin{subfigure}[b]{1\textwidth}
\includegraphics[width=1\textwidth,height=.3\textwidth]{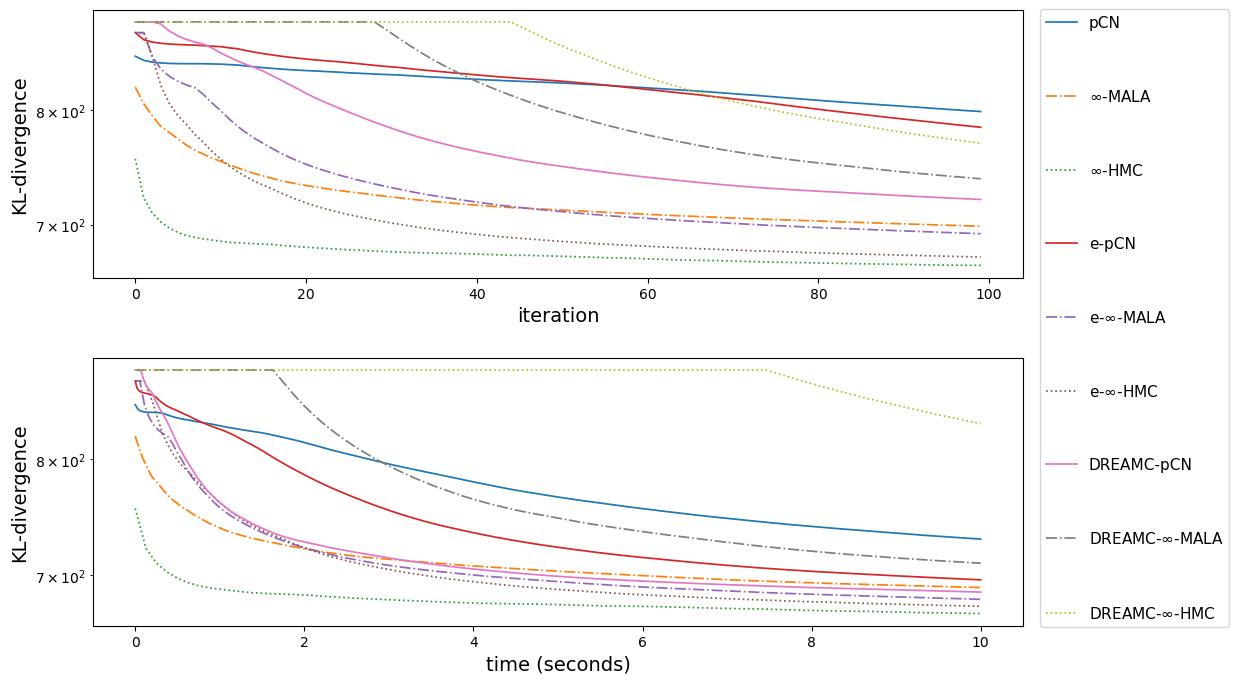}
\caption{The KL divergence between the posterior and the prior as a function of iteration (upper) and time (lower) respectively.}
\label{fig:adif_KLt}
\end{subfigure}
\vspace{-20pt}
\caption{Advection-diffusion inverse problem.}
\end{figure}

For the advection-diffusion inverse problem (Section \ref{sec:adif}), Figure \ref{fig:spatiotemporal_obs} plots 4 snapshots of these observations at 80 locations along the inner boundary. 
Figure \ref{fig:adif_acf} verifies DREAMC $\infty$-HMC is the most efficient MCMC algorithm that has the smallest autocorrelation as shown in the right panel. It is followed by other HMC algorithms and DREAMC $\infty$-MALA, which is even better than $\infty$-HMC.
Figure \ref{fig:adif_KLt} plots the KL divergence between the posterior and the prior in terms of iteration (top panel) and time (panel) respectively. As we can see, $\infty$-HMC converges the fastest.

\begin{figure}[t]
\centering
\includegraphics[height=.3\textwidth,width=1\textwidth]{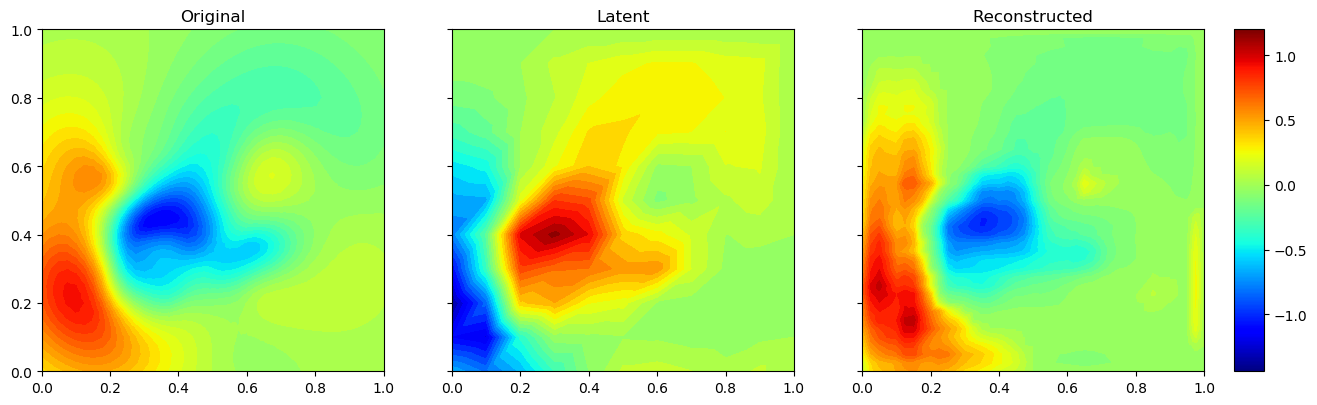}
\caption{Elliptic inverse problem: CAE compressing the original function (left) into latent space (middle) and reconstructing it in the original space (right).}
\label{fig:elliptic_reconstruction_cae}
\end{figure}

\begin{figure}[ht]
  \begin{center}
     \includegraphics[width=.495\textwidth,height=.3\textwidth]{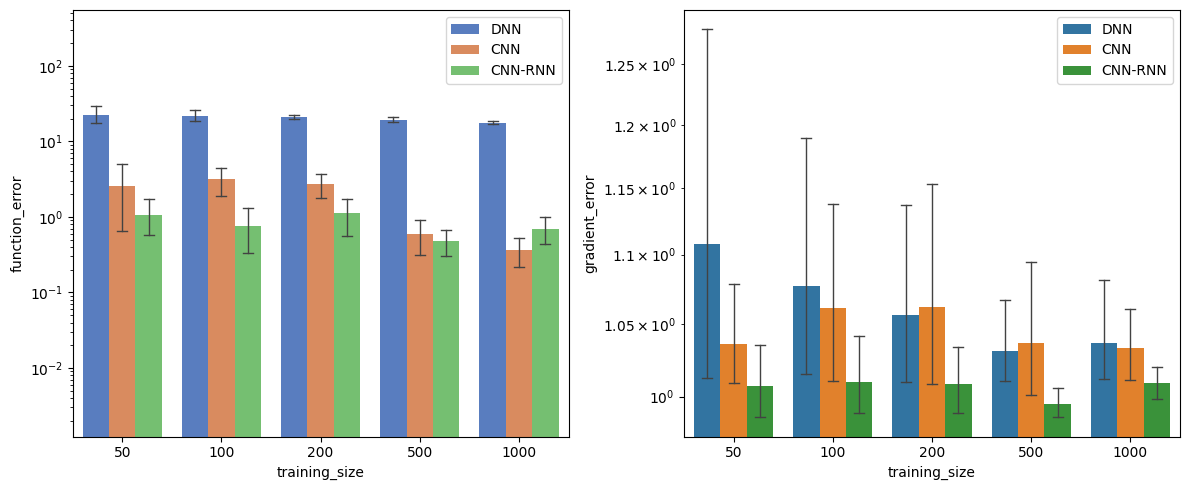}
     \includegraphics[width=.495\textwidth,height=.3\textwidth]{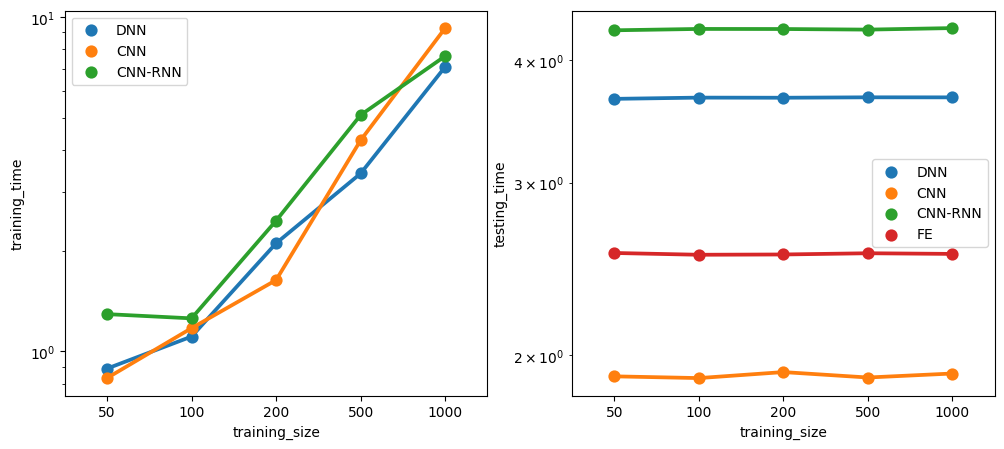}
  \end{center}
  \caption{Comparing the emulation $\mG^e: \mbR^{3413}\to\mbR^{1280}$ in an advection-diffusion inverse problem (Section \ref{sec:adif}) by DNN, CNN and CNN-RNN in terms of error (left) and time (right).
  Time is also compared with exact calculation of gradients (labeled `FE') using adjoint codes in testing.}
  \label{fig:cnn_cnnrnn}
\end{figure}

For spatiotemporal inverse problems such as advection-diffusion (Section \ref{sec:adif}), Figure \ref{fig:cnn_cnnrnn} suggests a combination of CNN (for spatial inputs) and RNN (temporal outputs) might perform better than CNN alone because RNN fits the time series better.
The left two panels show CNN-RNN yields the smallest function and gradient errors across different training sets. The right two panels show the computational cost is about the same as other NNs in training. Though it is less efficient in prediction in this example, CNN-RNN seems promising for efficient emulation in inverse problems with spatiotemporal data.

\bibliographystyle{siamplain}
\bibliography{references}

%% file: info_shared.tex
\usepackage{lipsum}
\usepackage{amsfonts}
\usepackage{graphicx}
\usepackage{epstopdf}
\usepackage{algorithmic}
\ifpdf
  \DeclareGraphicsExtensions{.eps,.pdf,.png,.jpg}
\else
  \DeclareGraphicsExtensions{.eps}
\fi

\usepackage{enumitem}
\setlist[enumerate]{leftmargin=.5in}
\setlist[itemize]{leftmargin=.5in}

\newsiamremark{remark}{Remark}
\newsiamremark{hypothesis}{Hypothesis}
\crefname{hypothesis}{Hypothesis}{Hypotheses}
\newsiamthm{claim}{Claim}

\headers{Bayesian UQ using Deep Neural Networks}{S. Lan, S. Li, and B. Shahbaba}

\title{Scaling Up Bayesian Uncertainty Quantification for Inverse Problems using Deep Neural Networks\thanks{Submitted to the editors DATE.
\funding{SL is supported by NSF grant DMS-2134256. BS is supported by NSF grant DMS-1936833 and NIH grant R01-MH115697.}}}

\author{Shiwei Lan\thanks{School of Mathematical and Statistical Sciences, Arizona State University, AZ
  (\email{slan@asu.edu}, \url{https://math.la.asu.edu/\string~slan/}).}
\and Shuyi Li\footnotemark[2]
\and Babak Shahbaba\thanks{Department of Statistics, University of California, Irvine, CA}}

\usepackage{amsopn}

\usepackage{graphicx}
\usepackage{amssymb}
\usepackage{amsmath}
\usepackage{amsfonts}
\usepackage{amssymb}
\usepackage{mathtools}
\usepackage{thmtools,thm-restate}
\usepackage{comment}
\usepackage{ifpdf}
\usepackage{bm}
\usepackage{xcolor}
\usepackage{scalerel}
\usepackage{stackengine,wasysym}
\usepackage{xr-hyper}
\usepackage{hyperref}
\usepackage[stable]{footmisc}
\usepackage{tikz-cd}
\usetikzlibrary{arrows,positioning}
\tikzset{
  shift left/.style ={commutative diagrams/shift left={#1}},
  shift right/.style={commutative diagrams/shift right={#1}}
}
\usepackage{subcaption}

\newtheorem{rk}{Remark}

\newcommand{\tp}[1]{{#1}^{\mathsf T}}
\renewcommand{\bar}{\overline}
\newcommand{\eps}{\epsilon}
\newcommand{\pa}{\partial}

\renewcommand{\eps}{\varepsilon}
\renewcommand{\epsilon}{\varepsilon}
\renewcommand{\Sigma}{\varSigma}

\newcommand{\tr}{\mathrm{tr}}

\DeclareMathAlphabet\mathbfcal{OMS}{cmsy}{b}{n}
\newcommand{\half}{\frac12}

\newcommand{\bv}{{\bf v}}

\newcommand{\bx}{{\bf x}}

\newcommand{\mC}{{\mathcal C}}
\newcommand{\mG}{{\mathcal G}}
\newcommand{\mK}{\mathcal{K}}

\newcommand{\mI}{{\mathcal I}}

\newcommand{\mN}{{\mathcal N}}

\newcommand{\mbX}{{\mathbb X}}
\newcommand{\mbY}{{\mathbb Y}}

\newcommand{\mbR}{{\mathbb R}}

\ifpdf
   \graphicspath{{./figure/PNG/}{./figure/PDF/}{./figure/}}
\else
   \graphicspath{{./figure/EPS/}{./figure/}}
\fi